\documentclass[titlepage]{amsart}

\usepackage[foot]{amsaddr}
\title{H2OPUS-TLR: High Performance Tile Low Rank Symmetric Factorizations using Adaptive Randomized Approximation}
\author{Wajih Boukaram}
\author{Stefano Zampini}
\author{George Turkiyyah}
\author{David Keyes}
\address{King Abdullah University of Science and Technology, Extreme Computing Research Center, Thuwal, Saudi Arabia}

\usepackage{graphicx}
\usepackage{amsmath,amssymb,mathtools,latexsym}
\usepackage{algorithmicx}
\usepackage{algorithm}
\usepackage[noend]{algpseudocode}
\usepackage{algpseudocode}
\algrenewcommand\alglinenumber[1]{\footnotesize #1}
\algrenewcommand\algorithmicindent{0.9em}%
\algblockdefx{ForAllp}{EndForAllp}[1] {\textbf{in parallel for} #1}%

\makeatletter
\ifthenelse{\equal{\ALG@noend}{t}}%
  {\algtext*{EndForAllp}}
  {}%
\makeatother

\algblock{Input}{EndInput}
\algnotext{EndInput}
\algblock{Output}{EndOutput}
\algnotext{EndOutput}
\newcommand{\Desc}[2]{\State \makebox[2em][l]{#1}#2}

\usepackage{array}     
\usepackage{url,tikz,pgfplots,xcolor}
\usetikzlibrary{spy,calc}
\usepackage{lipsum,adjustbox}

\usepackage{subcaption}
\usepackage[font={small,singlespacing}]{caption}
\newcommand{\batch}[1]{{#1_{\scriptscriptstyle \vert\kern-0.24ex\vert\kern-0.24ex\vert}}}

\begin{document}
\newcommand{\szcomment}[1]{\textcolor{red}{(#1)}}


\renewcommand{\shortauthors}{\footnotesize Boukaram, Zampini, Turkiyyah, Keyes}
\renewcommand{\shorttitle}{\footnotesize TLR Cholesky}

\begin{abstract}
\vspace*{6pt}
Tile low rank (TLR) representations of dense matrices partition them into blocks of roughly uniform size, where each
off-diagonal tile is compressed and stored as its own low rank factorization. They offer an attractive representation for many data-sparse dense operators that appear in practical applications, where substantial compression and a much smaller memory footprint can be achieved. TLR matrices are a compromise between the simplicity of a regular perfectly-strided data structure and the optimal complexity of the unbalanced trees of hierarchically low rank matrices, and provide a convenient performance-tuning parameter through their tile size that can be proportioned to take into account the cache size where the tiles reside in the memory hierarchy.

Despite their utility however, there are currently no high performance algorithms that can generate their Cholesky and $LDL^T$ factorizations and operate on them efficiently, particularly on GPUs. The difficulties in achieving high performance when factoring TLR matrices come from the expensive compression operations that must be performed during the factorization process and the adaptive rank distribution of the tiles that causes an irregular work pattern for the processing cores. In this work, we develop a dynamic batching operation and combine it with batched adaptive randomized approximations to remedy these difficulties and achieve high performance both on GPUs and CPUs. 

Our implementation  attains over 1.2 TFLOP/s in double precision on the V100 GPU, and is limited 
primarily by the underlying performance of batched GEMM operations. 
 The time-to-solution also shows substantial speedup compared to regular dense factorizations. The Cholesky factorization of covariance matrix of size $N = 131K$ arising in 2D or 3D spatial statistics, for example, can be factored to an accuracy $\epsilon=10^{-2}$ in just a few seconds. We believe the proposed GEMM-centric algorithm allows it to be readily ported to newer hardware such as the tensor cores that are optimized for small GEMM operations. 

\end{abstract}



\maketitle

\section{Introduction}

Cholesky and related LDL$^T$ factorizations of dense symmetric matrices are some of the most commonly utilized 
inner kernels in scientific and engineering simulations. 
Sampling from a multivariate normal distribution, 
operating on the fronts in a multi-frontal sparse solver, 
improving robustness of preconditioners, 
estimating maximum likelihood in Gaussian processes,  
solving with the Hessian of optimization problems, 
and regression in kernel-based machine learning, 
are just a few examples of embedding applications in which the factorization of a symmetric operator plays 
a key role in overall performance. LAPACK has long provided high quality CPU implementations of Cholesky 
and LDL$^T$ routines in different variants to account for different needs and matrix characteristics. 
Implementations of a subset of those routines are also available on GPUs in vendor-provided libraries. 
These routines provide the main workhorse for the needed factorizations when the matrix operands fit
comfortably in memory.

There are however two well-known bottlenecks of dense Cholesky and LDL$^T$ factorizations that limit their 
use. The first is memory. A memory footprint of $N^2$ floating point numbers becomes prohibitive to store in double precision on GPUs for $N$ larger than about a few tens of thousands, even on the recent large-memory 
GPUs. The other is the $\mathcal{O}(N^3)$ operations required to produce the Cholesky factor. Even with teraflop 
capabilities of modern GPUs, the cubic growth becomes overwhelming very quickly. There is a great need 
for reducing these resource intensive requirements both on CPUs and GPUs. The memory constraint is particularly 
problematic as modern computer architectures are generally evolving to have a decreasing ratio of memory 
capacity (and memory bandwidth) to processing power. 

One path for mitigating the growth in memory and arithmetic operations is to use a \emph{compressed} 
representation of the dense matrix by taking advantage of an underlying block low rank structure.
For a wide class of formally dense linear operators that arise in applications, including for example those mentioned earlier, the matrices are actually ``data sparse''. They require far less than the apparent 
$N^2$ values for their representation to high accuracy. This observation was initially made in the seminal 
work of Hackbusch \cite{hackbush99} in the context of integral equations which led to the development 
of hierarchical $\mathcal{H}$ and $\mathcal{H}^2$ matrices.  Hierarchical matrices exploit the block 
low rank structure \emph{hierarchically}, and in fact result in a memory footprint of $\mathcal{O}(N 
\log N)$ and asymptotically optimal $\mathcal{O}(N)$ in the case of $\mathcal{H}^2$ matrices with their 
hierarchical bases. 
\begin{figure}
  \begin{center}
      \includegraphics[width=.8\textwidth]{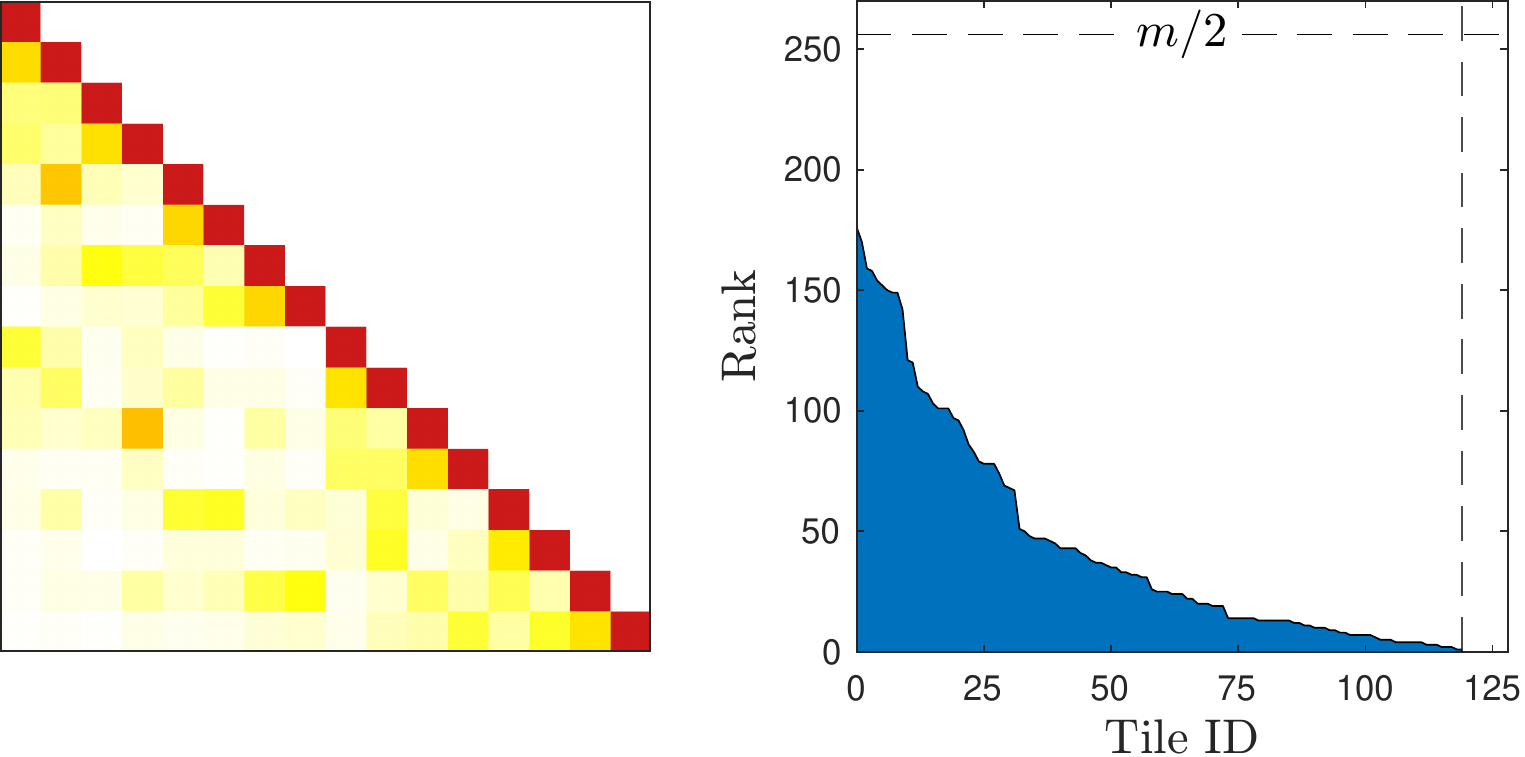}
      \caption{Illustrative TLR matrix of size 8K with tile size $m = 512$, and the rank distribution of its off-diagonal blocks. 
      Ratio of the shaded area and the rectangle bordered by the axes and dashed lines represents the amount of
      compression realized in the off-diagonal memory.}
      \label{fig:tlr}
  \end{center}
\end{figure}

Instead of using the full machinery of hierarchical matrices, the compression we consider in this paper is flat.
We decompose the matrix into a set of tiles, roughly uniform in size, and proportioned taking into account cache
sizes of the GPU/CPU processors. We should mention here that both the block low rank (BLR) and tile low rank (TLR)
nomenclature are used in the literature, roughly interchangeably. In this work, we will use the TLR terminology
which borrows from the ``tiled'' matrix data structures used in the multicore routines of PLASMA \cite{buttari09}
to expose fine grained parallelism. The TLR representation is also closely related to the block separable matrices described in \cite{gillman12}.

The implication of the tile low rank structure on the representation is that off-diagonal blocks $A_{ij}$, of size
$m \times m$, can be approximated to high accuracy by low rank splittings of the form $A_{ij} \approx U_{ij}
V_{ij}^T$ where $U_{ij}$ and $V_{ij}$ are $m\times k_{ij}$ with $k_{ij} \ll m$. The rank $k_{ij}$ naturally varies
between the blocks as dictated by their accuracy requirements. When $k_{ij} < m/2$, as is normally the case for a
data-sparse operator, the amount of storage for a block, $2 m k_{ij}$, is less than the $m^2$ storage required for
a dense representation. Figure \ref{fig:tlr} illustrates a TLR matrix structure obtained from a statistics
applications involving 8K points in a three-dimensional ball, and displays a distribution of the ranks of its off
diagonal tiles. In our representation, diagonal tiles, which normally have full rank, are stored in a dense
format, while the off diagonals are stored in the factored form $UV^T$. This can accommodate tiles that may be
full rank, or nearly so, at the cost of a slight memory consumption when $k_{ij} > m/2$. As our TLR representation
is rank-adaptive, the amount of compression realized by the compressed representation is directly related to the
rank distribution of the off-diagonal tiles, as illustrated in Figure \ref{fig:tlr}.



Factorization of a symmetric TLR matrix can in principle start with a block factorization algorithm, e.g., \cite[Sec 4.2.9]{golub13}, and operate on off-diagonal tiles using their low rank $U V^T$ representations. Unfortunately, this recipe is problematic from a performance viewpoint. The addition of low ranks blocks rapidly increases their apparent ranks, which then have to be truncated to the accuracy $\epsilon$ desired in the matrix factorization through a compression operation. Repeated tile recompressions are expensive and could wipe out many of the savings realized by performing the GEMM operations on tiles in the low rank representation. 

One of the challenges in performance-oriented Cholesky or LDL$^T$ algorithms is thus to minimize both the number of compressions that have to be performed, as well as the cost of the individual compressions. Many compression schemes are possible. SVD or rank revealing QR \cite{chan1987rank} are natural candidates to use. However they are expensive and require an explicit representation of the tile to compress. Randomized methods \cite{halko11} for constructing approximate decompositions can avoid accumulating the contents of a tile and yield better performance through their increased arithmetic intensity. The adaptive rank distribution of the tiles poses a challenge 
for efficient parallel execution of compression operations that we describe and address below. 

Another challenge in practice is that tile compression may result in loss of positive definiteness during a Cholesky factorization procedure of an original SPD matrix. The compression of an off-diagonal tile in effect introduces a perturbation in the matrix that must be carefully balanced to maintain definiteness and prevent the failure of the Cholesky factorization. We describe how Schur compensation can be used to remedy this problem, without incurring a performance penalty. Semi-definite matrices can also be accommodated by diagonal tile pivoting,
which is also done efficiently as it involves no data movement. 



Besides the performance and numerical algorithmic considerations, there is a practical requirement 
of performance portability in the development of high performance factorizations algorithms that target current and upcoming modern processors. The ability to efficiently utilize a variety of hardware architectures, including for example the tensor cores that are starting to be available on commercial processors, is obviously desirable.

\textbf{Contributions.} In this work, we propose and demonstrate high performance algorithms for symmetric factorizations in the TLR format. The contributions of the work are threefold.
\begin{itemize}
  \item \emph{Optimized tile compression.} Computations are organized so that (1) compression needs to be
  performed only once per tile of the final output factor $L$, and (2) tiles are compressed on the fly without
  generating dense representations of them first: the generator expression of a resulting tile is sampled via
  Adaptive Randomized Approximation (ARA) \cite{boukaram_hara_2019} to compute its low rank form \emph{ab initio}.

  \item \emph{Dynamic batched operations.} In order to efficiency utilize processing cores,
  especially on GPUs, compressions have to be batched. When using adaptive ranks, some tiles in the batch have
  larger ranks and require more ARA iterations to be compressed and must remain in processing, while the tiles
  that have converged must be removed, to either allow for new work to be fed to the processor or to allow
  processing resources to be used on the remaining unconverged tiles. We orchestrate this through the use of a
  parallel buffer that enables the parallelism to be controlled at a fine level through a dynamic update of the
  batch in progress.

  \item \emph{Performance portable GEMM-dominated algorithm.} As we show in the algorithm profile, up to 90\% of
  the runtime is spent in GEMM operations, which are the operations accelerated the most by current and
  next-generation hardware, allowing an automatic improvement in performance as newer hardware is used. 
  Cholesky performance attains more than 1.2 TFLOP/s in double precision on a V100
  NVIDIA GPU. The time-to-solution is sped up by two orders of magnitude compared to the dense factorization on
  representative matrices of size $N = 131K$ originating from applications in spatial statistics and 3D
  fractional diffusion. As an indication of absolute performance, a 2D spatial statistics covariance matrix of
  this size can be factored to a working accuracy of $\epsilon=10^{-2}$ in under $4$s on a V100 NVIDIA GPU, while a 
  covariance matrix of a 3D problem can be factored in under $14s$.
\end{itemize}
 
The resulting code is open-source and is included in the H2Opus library \cite{h2opus}. It provides implementations of positive definiteness-preserving Cholesky and $LDL^T$ factorizations. 

The rest of this paper is organized as follows. Section 2 surveys briefly related work on TLR and related
factorizations. Section 3 reviews some background related to adaptive randomized approximations used for
compression and the left-looking Cholesky decomposition that forms the basis of the proposed algorithm. Section 4
describes the core Cholesky algorithm. Section 5 describes enhancements to improve its robustness for insuring
positive definiteness, for treating ``near'' positive-definite matrices, for pivoting in the case of semidefinite
problems, and for generating $LDL^T$ factorizations. We show that these variations add little to the runtime complexity of the
algorithm. Section 6 presents representative performance results and we conclude with future directions in Section
7.

\section{Related Work}

Hierarchical matrix representations have long exploited the fact that off-diagonal blocks of the Schur complements of PDE discretizations
admit low rank approximations. Building on this observation, Amestoy \textit{et al.} \cite{amestoy15} proposed a block low rank representation of matrices, as an alternative to hierarchical matrix representations, in the context of multifrontal methods in sparse direct solvers. Their work shows that even though the TLR format is asymptotically less efficient than hierarchical approaches, it still delivers considerable gains both in terms of memory and flops reduction compared to dense representations. \cite{amestoy19a} presents a multithreaded TLR factorization for multifrontal solvers and analyzes its efficiency and scalability in shared-memory multicore environments. Algorithmic variants of the TLR factorization, including when the updates are compressed during the course of factorizations, are analyzed and tuned to overcome the challenges of using the TLR in multifrontal solvers. The simplicity and flexibility of the format also makes it easy to use as a general purpose algebraic solver in the MUMPS package \cite{mumps}.

The asymptotic complexity of TLR factorizations as they arise in factorization are analyzed in \cite{amestoy17}. With a bounded rank for tiles and few reasonable assumptions, a version of the algorithm that compresses the tiles as early as possible is shown to have an arithmetic complexity of $\mathcal{O}(N^2)$ and a memory complexity of $\mathcal{O}(N^{1.5})$ assuming an increasing tile size as $\mathcal{O}(N^{0.5})$. The constants in these estimates involve the block ranks $k_{ij}$, which are obviously affected by the accuracy, $\epsilon$, desired. For problems arising from elliptic boundary value problems in $d$ dimensions, the maximum rank grows as $\mathcal{O}(|\log \epsilon|^{d+1})$ \cite{bebendorf16}, and is used to proportion the level block sizes in $\mathcal{H}$ matrices. In TLR representations, many tiles will generally have small ranks and the use of adaptive ranks substantially improves the constants of the estimates. The use of Strassen's algorithm in the inner GEMMs can reduce the theoretical complexity of the factorization even further to $\mathcal{O}(N^{1.9})$ \cite{jeannerod19}.

TLR approximation of dense matrices is also now available in the matrix package STRUMPACK \cite{strumpack}, both in multithreaded and distributed memory versions. Various shared memory implementations are also offered with a sparse supernodal solver PaStiX \cite{pichon18}.
A Cholesky factorization algorithm in the TLR format for large scale distributed memory machines was presented in \cite{keyes20,cao20}. The algorithm exploits the capabilities of task-based dynamic runtime systems to distribute the workload arising from the directed acyclic graph encapsulating the computational and data flow of the blocked algorithm.
A multicore Cholesky factorization using a dynamic runtime system was also presented in \cite{salvana21} where it was used in Gaussian log-likelihood function evaluations involving covariance matrices.

While low rank factorizations have been mostly used as approximate direct solvers or as quality preconditioners in various settings, they have also shown their practicality in other use cases. In \cite{higham19} for example, the TLR $LU$ factorization is used as a first step towards the construction of a globally low rank approximation of the error $U^{-1}L^{-1}A - I$, which is then shown to result in more effective preconditioners for ill-conditioned problems. \cite{cao21a} uses a TLR factorization for computing high-dimensional multivariate normal and Student-t probabilities, and shows that the convergence rate of Monte Carlo methods for computing the large dimensional integrations can be substantially improved.

Ordering unknowns and equations has a marked effect on rank distributions of TLR representations of matrices and their factors. Optimal orderings for minimizing ranks are generally not known nor are practical. The heuristics used for grouping unknowns into tiles in low spatial dimensions ($d = 2$, $3$) are generally those developed for clustering in hierarchical matrices. KD-trees, with their mean and median variants for how the splitting are made \cite{boukaram19a}, are a practical and scalable construction whose leaves can then be used for the TLR blocking. Morton orderings and other space filling curves have also been used to generate tilings for matrices in low spatial dimensions \cite{genton18,cao21b}. In higher dimensional spaces, such as the feature spaces that appear in machine learning applications, approximate nearest neighbor \cite{chavez20,rebrova18,march15} are computed based on random projection trees. These are generalizations of KD-trees, where the direction of the median split is randomized and is not one of the coordinate dimensions.

Proposals to augment the basic TLR format with features from $\mathcal{H}$ and $\mathcal{H}^2$ matrix representations have been presented with the goal of reducing the asymptotic memory and arithmetic complexities of factorization. An $\ell$-level BLR where the tiles are themselves TLR matrices was suggested in \cite{amestoy19b}. A BLR$^2$ representation \cite{ashcraft21} where the tiles in a given block row or block column share the same bases ($A_{ij} = U_i S_{ij} V_j^T$) borrows from the $\mathcal{H}^2$ representation \cite{borm10}. 
The Lattice-$\mathcal{H}$ format where the diagonal blocks are stored as $\mathcal{H}$ or $HSS$ \cite{rouet16} matrices has also been suggested as a means for bridging the gap between TLR and optimal-complexity hierarchical matrices. \cite{chavez18} uses an ACR preconditioner, which is effectively a tridiagonal TLR matrix whose tiles are hierarchical matrices. 

Despite the interest in TLR representations and their obvious relevance to various applications, there are currently no GPU hosted algorithms that can factor and operate on them efficiently. The goal of this paper is to present high-performance algorithms that run efficiently on GPUs and CPUs, and that are ready to exploit the next generation hardware that is expected on the horizon.

\section{Background}

\subsection{Adaptive Randomized Approximation}
\label{sec:ara}
A core low rank linear algebra operation is compression, where the rank of a matrix is reduced to 
satisfy a specific rank or error threshold. For example, applying a low rank update $U_l 
V_l^T$ to a tile in a TLR matrix $A(i,j) = U(i,j)V(i,j)^T$ is accomplished simply by appending the columns of $U_l$ to $U(i,j)$ and the columns of $V_l$ to $V(i,j)$, increasing the rank of the tile. To keep memory consumption low, this rank should be reduced by compression to a specific rank or accuracy threshold.
Any compression kernel produces a rank $k$ approximation of an $n 
\times m$ matrix $A$ in the form $UV^T$, where $U$ and $V$ are $n \times k$ and $m \times k$ matrices. 
Many algorithms can produce these approximations, such as the singular value decomposition (SVD), adaptive 
cross approximation (ACA) \cite{rjasanow2002adaptive}, rank revealing QR decomposition (RRQR) \cite{chan1987rank}, and
interpolative decomposition (ID) \cite{cheng_2005} to cite a few. The SVD algorithm produces the smallest rank 
possible for a given error threshold, but it is expensive to compute. Faster algorithms 
like ACA, RRQR or ID require direct access to the entries of a matrix.
On the other hand, randomization methods \cite{halko11} can generate an approximation using only black box matrix 
vector products without accessing matrix entries.
The most basic method generates a fixed rank $k$ approximation by taking the product $Y = A \Omega$, 
where $\Omega$ is a set of $k$ random vectors, and orthogonalizing $Y$ to produce an approximate basis 
$Q$ for the columns of $A$. The matrix is then projected into the basis to produce the right low rank 
factor $B=A^T Q$, giving us the required approximation of $A \approx Q B^T$. While this operation is 
rich in Level 3 BLAS, the rank of the approximation of a matrix that satisfies an error threshold is 
rarely known beforehand. Adaptive methods that sample the matrix one vector at a time were developed 
to address this limitation, giving up the performance benefits of BLAS 3. To reclaim those benefits, 
and in the quest for performance portability, we will make use of block algorithms that sample blocks 
of vectors at a time. 
Algorithm \ref{alg:ara} shows a high level summary of the Adaptive Randomized Approximation (ARA) method; 
the matrix is sampled in blocks of $bs$ vectors at a time, iteratively constructing the orthogonal basis $Q$ until 
the convergence threshold $QB^T \leq \epsilon$ is satisfied. The \texttt{orthog} routine uses two iterations of block 
Gram Schmidt orthogonalization where the QR factorization of each panel is implemented using mixed precision Cholesky QR.
We refer the interested reader to \cite{boukaram_hara_2019} for a more detailed description of a batched ARA algorithm designed for GPU execution.

\begin{algorithm}[H]
\caption{Adaptive Randomized Approximation}
\label{alg:ara}
\begin{algorithmic}
\Procedure{ARA}{$A$, $bs$, $\epsilon$}
	\State $e = 1$
	\State $n = $ size$(A)$
	\State $Q = []$
	\While{$e > \epsilon$} 
		\State $\Omega = $ randn $(n, bs)$ \Comment{Generate random vectors from a normal distribution}
		\State $Y = A \Omega$	\Comment{Sample $A$ using matrix vector products}
		\State $[Y, R] = $ orthog $(Q, Y)$ \Comment{Make $Y$ orthogonal to $Q$}
		\State $Q = [Q, Y]$ \Comment{Append the orthogonal columns to the basis}
		\State $e = $ convergence $(R)$ \Comment{Check for convergence}
	\EndWhile
\EndProcedure
\end{algorithmic}
\end{algorithm}

\subsection{Dense Cholesky Decomposition}

Given an $N \times N$ real symmetric positive definite matrix $A$, the Cholesky decomposition computes 
a factorization of the matrix as the product of a lower triangular matrix $L$ and its transpose $A = 
L L^T$. At every step $k$ of the algorithm, the diagonal element is inverted, the column $k$ 
of $A$ is transformed into a column of $L$ and the remaining $(N-k) \times (N-k)$ submatrix of $A$ is updated 
using the symmetric rank-1 update defined by the column $k$ of $L$. Tiled variants of the algorithm that 
partition the rows and columns of the matrix into tiles were developed to leverage the more arithmetically 
intensive Level 3 BLAS operations and to expose greater parallelism \cite{buttari09}.

\begin{figure}
     \centering
     \begin{subfigure}[b]{0.45\textwidth}
         \centering
         \includegraphics[width=\textwidth]{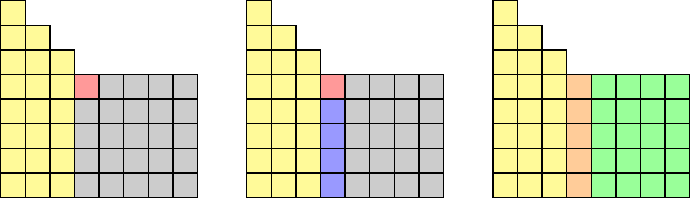}
         \caption{Right looking tiled Cholesky}
         \label{fig:right_tiled_chol}
     \end{subfigure}
     \hfill
     \begin{subfigure}[b]{0.45\textwidth}
         \centering
         \includegraphics[width=\textwidth]{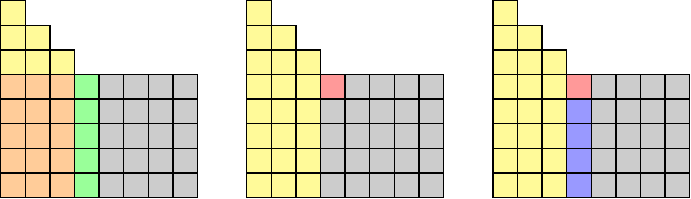}
         \caption{Left looking tiled Cholesky}
         \label{fig:left_tiled_chol}
     \end{subfigure}
     \caption{The three major phases of each step of the left and right looking variants of tiled Cholesky 
decomposition. The tiles of the trailing submatrix of $A$ are shown in gray, the processed tiles of $L$ 
are in yellow, diagonal tile Cholesky is in red, triangular solves are in blue, updated blocks are in 
green and the source of the updates are in orange.}
     \label{fig:tiled_chol}
\end{figure}

In the tiled algorithm, the diagonal solve of a scalar becomes an unblocked Cholesky solve of the diagonal 
tile, the transformation of the single column becomes a triangular solve of the sub-diagonal tiles of 
the block column, and the rank-1 update is transformed into a rank-$b$ update of the trailing submatrix, 
where $b$ is the tile size. Algorithm \ref{alg:right_cholesky} shows the tiled version of the Cholesky 
decomposition, where $A(i, j)$ refers to the $b \times b$ tile in the $i$-th block row and $j$-th block 
column. This variant is known as the right looking Cholesky, since at each step $k$, the low rank updates 
are applied to the tiles to the right of the block column $k$ as shown in Figure \ref{fig:right_tiled_chol}. 
The left looking variant applies all the updates from the left of the column before carrying out the 
diagonal and triangular solves on the column tiles as shown in Algorithm \ref{alg:left_cholesky} and 
Figure \ref{fig:left_tiled_chol}. When implemented using dense linear algebra, each version has its
benefits, such as good load balancing for right looking and reduced I/O operations for left looking. 
With low rank linear algebra, the left looking variant has more attractive properties when coupled with a compression scheme like ARA as discussed in the next section. 



\begin{minipage}{0.5\textwidth}
\begin{algorithm}[H]
\caption{Right Looking Cholesky}
\label{alg:right_cholesky}
\begin{algorithmic}[1]
\Procedure{rchol}{$A, b$}
	\State $nb$ = size$(A) / b$
	\For{$k = 1 \rightarrow nb$} 
		\State $L(k, k) = $ chol$(A(k, k))$
		\For{$i = k+1 \rightarrow nb$}    
			\State $L(i, k) = A(i, k) / L(k, k)^T$
		\EndFor
		\For{$j = k+1 \rightarrow nb$}
			\For{$i = j \rightarrow nb$}    
				\State $A(i, j) = A(i, j) - L(i, k) L(j, k)^T$
			\EndFor
		\EndFor
	\EndFor
\EndProcedure
\end{algorithmic}
\end{algorithm}
\end{minipage}
\hfill
\begin{minipage}{0.48\textwidth}
\begin{algorithm}[H]
\caption{Left Looking Cholesky}
\label{alg:left_cholesky}
\begin{algorithmic}
\Procedure{lchol}{$A, b$}
	\State $nb$ = size$(A) / b$
	\For{$k = 1 \rightarrow nb$} 
		\For{$j = 1 \rightarrow k-1$}		\label{alg:left_cholesky:lru_start}
			\For{$i = k \rightarrow nb$}
				\State $A(i, k) = A(i, k) - L(i, j) L(k, j)^T$
			\EndFor                           \label{alg:left_cholesky:lru_end}
		\EndFor
		\State $L(k, k) = $ chol$(A(k, k))$
		\For{$i = k+1 \rightarrow nb$}    
			\State $L(i, k) = A(i, k) / L(k, k)^T$
		\EndFor
	\EndFor
\EndProcedure
\end{algorithmic}
\end{algorithm}
\end{minipage}

\section{Tile Low Rank Cholesky}
\label{sec:tlr_potrf}
The right looking Cholesky applies the low rank updates to the trailing submatrix of $A$ as soon as the 
column of $L$ has been determined. Updating a low rank tile should be then followed by compression to keep the memory consumption of the matrix tiles low. Repeatedly updating and compressing the tiles after each update will incur large computational costs, with the vast majority 
of computations performed using relatively low efficiency kernels. On the other hand, the left looking variant 
updates each block only once using all of the low rank updates to its left. While naively applying each 
low rank update one at a time will severely limiting parallelism, it would appear that a parallel reduction style low rank update, where pairs of low rank updates are successively summed up, offers a solution. 

However, such a reduction greatly increases memory consumption 
and introduces the problem of determining the optimal order of updates to avoid rank growth in the intermediate 
operands. To overcome these issues, we treat the final updated block as a matrix expression that can 
be sampled and compressed using ARA in a single compression step using highly efficient kernels. 

In this section, we start by discussing the details of the required building blocks for
the left looking TLR Cholesky algorithm, including the core sampling process and its use in an optimized batched ARA. We then present the overall factorization and show how the vector products and triangular solves can efficiently use the computed factors.

\subsection{Left looking sampling}

The left looking low rank updates to a block $A(i, k)$ at lines \ref{alg:left_cholesky:lru_start}-\ref{alg:left_cholesky:lru_end} 
of Algorithm \ref{alg:left_cholesky} updates each block in the column $k$ as
\begin{equation}
\label{eq:left_lru}
A(i, k)= A(i, k) - \sum_{j=1}^{k-1} L(i, j) L(k, j)^T = A(i, k) - \sum_{j=1}^{k-1} U(i, j)V(i,j)^T V(k, 
j) U(k,j)^T ,
\end{equation}
where $U$ and $V$ are the low rank factors of the tiles of $L$.
Within the ARA algorithm, each operand in the sum can be efficiently sampled by computing four intermediate 
dense matrix-matrix products as
\begin{equation}
\label{eq:sample_lru}
Y_j = U(i, j) \left(V(i,j)^T \left( V(k, j) \left( U(k,j)^T \Omega_j \right) \right) \right).
\end{equation}

\begin{algorithm}[b]
\caption{Left Looking Sampling}
\label{alg:left_looking_sampling}
\begin{algorithmic}[1]
\Procedure{sampleLeft}{$A, k, ri, ws, \Omega$}
	\Input
	\Desc{$A$}{TLR Matrix}
	\Desc{$k$}{Current panel index}
	\Desc{$ri$}{Row indices sampling tiles}
	\Desc{$ws$}{Workspace}
	\Desc{$\Omega$}{Random sampling vectors}
	\EndInput
	
	\Output 
	\Desc{$Y$}{Samples for all the tiles}
	\EndOutput
	
	\State $pb = $ buffers $(ws) / $ size$(ri)$  \label{alg:left_looking_sampling:parallel_buffers} \Comment{Number 
of columns to sample in parallel}
	\State $ps = pb / (k - 1)$           \Comment{Number of sampling steps}
	\For{$j = 1 \rightarrow ps$} 
		\State $\left[U_{kj}, V_{kj}, U_{ij}, V_{ij}\right] = $ marshalTiles $(A, ri, j, pb, k)$ \Comment{Marshal 
tile data}
		\State $\left[ W_1, W_2, W_3, W_4 \right] = $ marshalWorkspace $(ws, pb)$ \Comment{Get sampling buffers}

		\State $W_1 = $ batchGemm $(U_{kj}^T, \Omega)$  \label{alg:left_looking_sampling:gemm_start}
		\State $W_2 = $ batchGemm $(V_{kj}, W_1)$
		\State $W_3 = $ batchGemm $(V_{ij}^T, W_2)$
		\State $W_4 = W_4 + $ batchGemm $(U_{ij}, W_3)$ \label{alg:left_looking_sampling:gemm_end} \Comment{$W_4$ 
accumulates the samples}
	\EndFor
	\State $Y = $ reduceBuffers $(W_4)$ \Comment{Reduce all parallel buffers into $Y$}
	\State $Y = $ sampleColumn $(A, k, ws, \Omega) - Y$  \Comment{Column samples and update of $Y$} \label{alg:left_looking_sampling:sample_col}

\EndProcedure
\end{algorithmic}
\end{algorithm}

Algorithm \ref{alg:left_looking_sampling} describes the process of sampling the sum in Equation \ref{eq:left_lru} 
for a column $k$ assuming that the matrix $A$ has been partially overwritten by the triangular factor $L$.
To increase parallelism we sample multiple updates into independent buffers $Y_j$ that get finally reduced 
to assemble the final samples $Y$ needed by ARA.
This leads to a tradeoff that must be made between absolute performances and memory consumption for the 
parallel matrix buffers.
In our implementation, we first generate batches of matrix-matrix products by marshaling the necessary 
data (from the TLR matrix and our workspace) and then execute the products using non-uniform batched 
\texttt{GEMM} routines. Marshaling is a lightweight process, 
involving pointer operations and no data movement, needed to allow batched executions.

Since we would like to keep occupancy high by producing large batches, the number of parallel matrix buffers $Y_j$ for each block 
is dynamically determined to be as large as the allocated workspace allows (line \ref{alg:left_looking_sampling:parallel_buffers} 
of Algorithm \ref{alg:left_looking_sampling}). While this implies that more time is spent in the final 
parallel reduction operation, the increased occupancy leads to faster solution times.
After the sum of the low rank updates has been sampled, the final step (line \ref{alg:left_looking_sampling:sample_col}) 
is to sample the original block to finalize the sampling of the matrix expression in Equation \ref{eq:left_lru}.

\begin{figure}[h]
  \begin{center}
      \includegraphics[width=.6\textwidth]{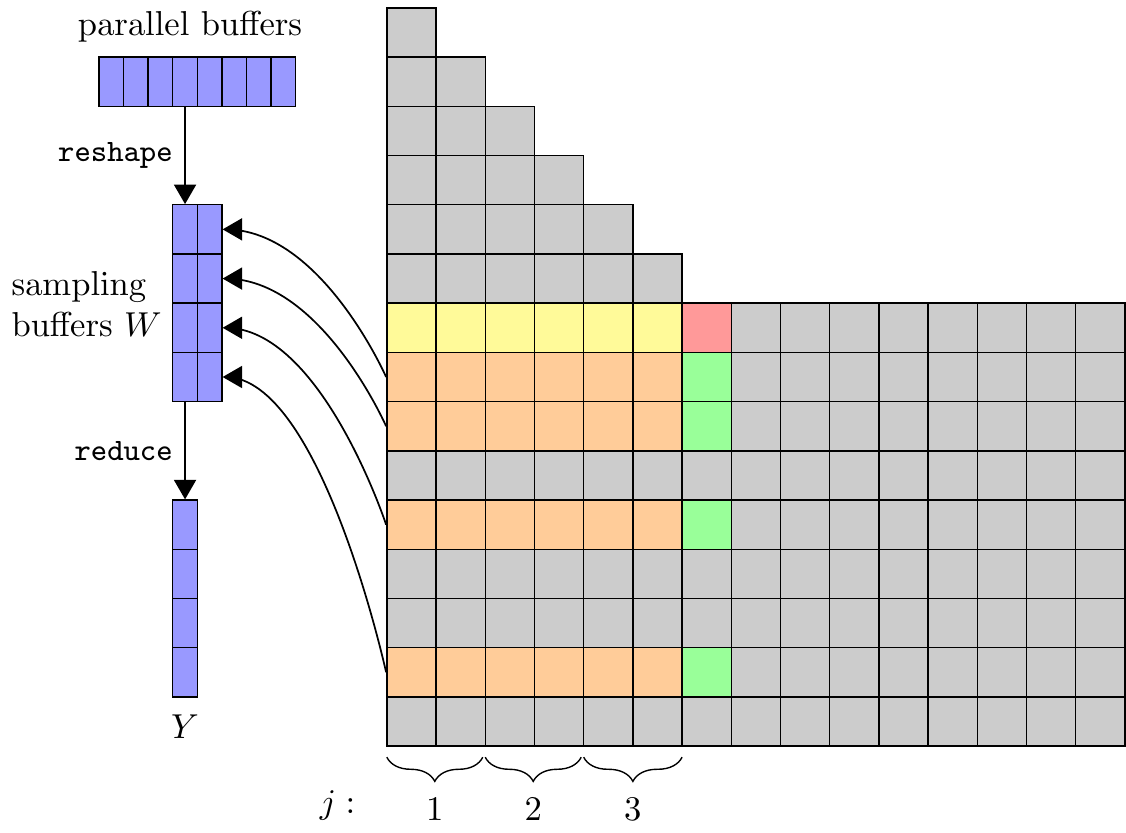}
      \caption{Parallel sampling of low rank updates to the tiles in green. Eight buffers allocated in the workspace are reshaped into a $4 \times 2$ matrix of buffers that sample the six tile updates in three serial steps. The final samples $Y$ are produced by a parallel row reduction of the buffer matrix.}
      \label{fig:left_looking_sampling}
  \end{center}
\end{figure}

Figure \ref{fig:left_looking_sampling} shows this process at step $k=7$ for sampling a set of four of the updated tiles in green with the source of the updates in orange and yellow. The allocated parallel buffer of eight tiles in blue can be used to sample two of the updates of Equation \ref{eq:sample_lru} to those four tiles in parallel. This leads to three total serial steps for sampling the six set of updates to the left of the tiles that need to be updated. Increasing the size of the parallel buffers of the workspace to twelve tiles would allow sampling three updates per tile in parallel and reduce the serial steps to two, increasing parallelism at the cost of memory. The final step is a parallel row reduction of the buffers to produce the output of the sampling $Y$.

\subsection{Optimized Batched ARA}
Algorithm \ref{alg:tlr_ara} makes use of the sampling routine of Algorithm \ref{alg:left_looking_sampling}
and compresses the updated tiles in a column 
$k$.
A naive implementation may simply marshal all tiles within the column that is being updated within the 
same batch; however, the ranks of the tiles can vary significantly within the same column. For example, 
the majority of the tile ranks from statistics applications are quite small, with only a few outliers 
having significantly larger ranks than the others. If we sample all blocks within the column at the same 
time, the tiles with smaller ranks will converge very quickly, leaving a batch of just a few tiles. In 
this case, the processor will be oversubscribed with work for a short time and then starved for a long 
time.

In order to cope with general rank distributions within the same block column and the adaptive nature 
of the ARA algorithm, we initially sort the tiles by their original rank in descending order, since 
a tile with a large rank in $A$ tends to keep a large rank in its triangular factor $L$.
We then marshal only a subset of the tiles in the column and dynamically add the remaining ones to the 
batch as the tiles in the subset converge. This naturally allows tiles
of high rank to stay in the processing batch as long as they need. 
For example, Figure \ref{fig:left_looking_sampling} shows the sampling process for a subset of four updated tiles in green. If two tiles converge, they are removed from the subset and the next two of the remaining four grey tiles in the same column are added to the subset, along with the unconverged ones. 

As the size of the subset decreases, whether that is due to 
tiles converging or to the progress of the overall factorization, the batch size naturally decreases 
as well. 
A more advanced scheduler could start sampling blocks in the next column to alleviate
this burden and increase occupancy, and it would be an interesting direction to take the algorithm in the future.

The routine \texttt{getConvergedTiles} determines the indices within the current subset $ri$ that have 
converged based on the error threshold $\epsilon$ and the routine \texttt{updateSubset} replaces the 
converged tiles with tiles from the remaining tiles in the full set of row indices $r$.
When $ri$ is empty, all tiles have converged and we can proceed with the projection of the updated tiles 
into the approximate basis of the tiles $Q$. The routine \texttt{sampleLeftT} is almost identical to 
\texttt{sampleLeft} but samples the transpose of the low rank updates and the column tiles. 

\begin{algorithm}[t]
\caption{Left Looking Cholesky TLR ARA Update}
\label{alg:tlr_ara}
\begin{algorithmic}[1]
\Procedure{cholARAUpdate}{$A, bs, k, \epsilon, ws$}
	\Input
	\Desc{$A$}{TLR Matrix}
	\Desc{$bs$}{Block samples for ARA}
	\Desc{$k$}{Current panel index}
	\Desc{$ws$}{Workspace}
	\EndInput
	
	\Output 
	\Desc{$Q$}{Approximate basis for the updated tiles}
	\Desc{$B$}{Right low rank factor of the approximated updated tiles}
	\EndOutput
	
	\State $Q = []$
	\State $b = $ blocksize $(A)$
	\State $r = $ sortRanks $(A, k)$  \Comment{Sort the column tile ranks in descending order}
	\State $ri = $ subset $(r)$ 	  \Comment{Take a subset of column tiles}
	\While{$ri \neq \varnothing$}
		\State $\Omega = $ batchRandn $(b, bs, $ size $(ri))$
		\State $Y = $ sampleLeft$(A, k, ri, ws, \Omega)$ \Comment{Sample the low rank updates}
		\State $[Y, R] = $ batchOrthog $(Q, Y)$  \Comment{Make Y orthogonal to Q}
		\State $Q = [Q, Y]$ 				\Comment{Append the orthogonal columns to the basis}
		\State $ci = $ getConvergedTiles $(R, \epsilon)$ \Comment{Grab the converged tile indices}
		\State $ri = $ updateSubset $(ri, ci, r)$ \Comment{Remove converged. Add work from remainder}
	\EndWhile 
	\State $B = $ sampleLeftT$(A, k, ri, 1, k, ws, Q)$ \Comment{Project into the approximate basis}
\EndProcedure
\end{algorithmic}
\end{algorithm}

\subsection{TLR Cholesky}
Using the routines from the previous sections, we can now build the complete left looking TLR Cholesky 
in Algorithm \ref{alg:tlr_chol}. The low rank updates that are applied to the dense diagonal tile are 
expanded into their dense form and added as dense matrices to the diagonal tile before it is factorized 
using dense Cholesky decomposition.
This part of the computation can be in principle be overlapped with the rest of the
computation to increase occupancy on the GPU. However, as we further discuss in Section 6, 
the currently available non-uniform batched matrix-matrix product kernels 
are not yet fully asynchronous to allow for this overlap.

The left looking TLR ARA algorithm then computes an approximation
$QB^T$ for each of the updated tiles. The final step is a triangular solve on the updated tiles with 
the factorized diagonal tile: $A(i,k) = A(i, k) / L(k, k)^T = U(i, k) V(i, k)^T / L(k, k)^T$. This boils 
down to solving for the new right low rank factor of each block as $V(i, k) = L(k, k) / V(i, k)$ using 
a batched non-uniform \texttt{trsm} routine. 

\begin{algorithm}[t]
\caption{Left Looking TLR Cholesky}
\label{alg:tlr_chol}
\begin{algorithmic}[1]
\Procedure{TLRChol}{$A, bs, \epsilon, ws$}
	\Input
	\Desc{$A$}{TLR Matrix}
	\Desc{$bs$}{Block samples for ARA}
	\Desc{$ws$}{Workspace}
	\EndInput
	
	\Output 
	\Desc{$A$}{Lower triangular matrix $L$ overwriting $A$}
	\EndOutput
	
	\State $nb = $ blocks$(A)$
	\For{$k = 1 \rightarrow nb$}
		\State $A(k, k) = $ updateDense $(A, k, ws)$ \Comment{Dense updates of diagonal block}
		\State $A(k, k) = $ chol $(A(k, k))$	     \Comment{Dense Cholesky}
		\State $[Q, B] = $ cholARAUpdate $(A, bs, k, \epsilon, ws)$  \Comment{Approximate updated column tiles}

		\State $B = $ batchTrsm $(A(k, k), B, nb - k - 1)$ \Comment{Triangular solve}
		\State $A = $ updateTiles $(A, k, Q, B)$ \Comment{Replace old tiles with the updated ones}
	\EndFor
\EndProcedure
\end{algorithmic}
\end{algorithm}

\subsection{TLR Matrix Vector Product and Triangular Solves}
Algorithms for the tile low rank matrix vector products and triangular solves, 
needed to operate on the factorizations produced, follow the same general method that was used to implement the
previous TLR operations.  Computations that can be completed in parallel are first marshaled into batches followed by launching 
non-uniform batched matrix-matrix products and triangular solves. For the matrix-vector product, we increase parallelism by 
splitting the low rank block columns into independent sets of products stored in a set of output buffers followed by a reduction 
on the buffers into the final output. The low rank products with each block of the input vector is carried out as a sequence of two 
products with each of the low rank factors of the tiles. The tile operation in triangular solves are also marshaled such that each 
step $k$ updates the output in parallel after the diagonal block solve. For example, for the lower triangular solve, the block $x(k)$ 
of the input is updated with a dense triangular solve with the diagonal block $L(k, k)$ followed by a parallel update of all blocks 
$x(i) = L(i, k) x(k) = U(i,k) \left(V(i,k)^T x(k)\right)$ for $i=k+1\rightarrow nb$ as shown in Algorithm \ref{alg:tlr_trsm}. 

\begin{algorithm}[b]
\caption{TLR Triangular Solve}
\label{alg:tlr_trsm}
\begin{algorithmic}[1]
\Procedure{TLRTRSM}{$L, y$}
	\Input
	\Desc{$L$}{Lower triangular TLR Matrix}
	\Desc{$y$}{Input vector}
	\EndInput
	
	\Output 
	\Desc{$x$}{Solution of the system $L x = y$}
	\EndOutput
	
	\State $nb = $ blocks$(A)$
	\State $x = y$
	\For{$k = 1 \rightarrow nb$}
		\State $x(k) = $ trsm $(L(k,k), x(k))$ \Comment{Dense triangular solve}
		\State $[X, U, V] = $ marshalTRSM $(L, x, k)$ \Comment{Marshal the tile and input data}
		\State $T = $ batchGemm $(V^T, x(k))$
		\State $X = X - $ batchGemm $(U, T)$
	\EndFor
\EndProcedure
\end{algorithmic}
\end{algorithm}

\section{Extensions and LDL$^T$ factorization}
\label{sec:extensions}

In this section we describe three extensions to the core factorization algorithm of Section \ref{sec:tlr_potrf} to
expand the range of matrices that it can handle. Our first extension seeks to avoid the breakdown that may happen
as parts of matrix tiles are dropped in the process of compression. The second extension introduces symmetric
tile pivoting to handle semidefinite and ill-conditioned cases. The third extension produces an LDL$^T$ factorization
avoiding the Cholesky factorization of the diagonal blocks. Our primary goal is to show that these practical
extensions are readily incorporated in our algorithmic template and incur only a small runtime penalty. 


\subsection{Modifications for preserving definiteness}

Classical analysis of the stability of scalar dense Cholesky factorization \cite{golub13,wilkinson68} shows that
when $C u \kappa \le 1$, where $C$ is a size dependent factor, $u$ is the machine precision
used, and $\kappa$ is the condition number of the matrix, the Cholesky process runs to completion without
encountering negative pivots. While we do not have a formal analysis for TLR computations, intuition and practical experience with these algorithms suggest that
the threshold, $\epsilon$, we use in ARA compression effectively plays the role of the machine precision, and
therefore when $\epsilon \kappa$ is large enough, the Cholesky process may result in block diagonal tiles that fail to factor having lost positive definiteness after being updated. For ill-conditioned matrices, this may require undesirably tight compression thresholds to
be used or matrix modifications to preserve its positive definiteness.

The most common such modification is diagonal shifting, where a multiple of the identity $sI$ is added to $A$. 
When $s \le \epsilon$, the $A+sI$ perturbs the diagonal tiles by a quantity under the compression threshold and, as long the result is good enough as an approximate solver or as a preconditioner for the target application, this is a convenient strategy. However, more refined modifications, that use the tiles encountered in the factorization, are possible and we discuss below how we incorporate them.

\subsubsection{Schur Compensation}


Adding positive semidefinite terms to Schur complements in the Cholesky factorization is an idea that appears in many contexts. In incomplete sparse factorization, for instance, it is common that breakdown occurs when off-diagonal entries are discarded. A number of remedies for adding a compensating Schur modification in the form of positive semidefinite blocks have been proposed to deal with it \cite{scott14}. In the Cholesky factorization of hierarchical HSS matrices, a positive semidefinite compensation consisting of the difference between the Schur complement update and its compressed representation is (implicitly) added to obtain a positive definite result \cite{xia_ming_schur_compensation_2010}. Related methods for preserving definiteness are described in \cite{chow18}.

In a similar vein to \cite{xia_ming_schur_compensation_2010}, we propose a strategy where the error introduced by the compression is compensated for by adding positive semidefinite terms of the same order of the compression error when updating diagonal blocks. 
Instead of simply applying all the accumulated updates $D_k$ to the dense diagonal block $A(k, k)$, the update 
is first compressed to the threshold $\epsilon$ to produce $\bar{D}_k$ and the semipositive difference $D_k - \bar{D}_k$ is added as a compensating term to $A(k, k)$. We have also found it convenient to use diagonal compensation \cite{axelsson94} and add these terms to the diagonal elements as $\mathrm{rowsum} |D_k - \bar{D}_k|$. 
This is a simple modification to Algorithm \ref{alg:piv_tlr_chol}, converting line \ref{alg:piv_tlr_chol:update_2} 
to add the compensation that the update $D_k$ necessitates: 
$A(k, k) = A(k, k) - D_k + \texttt{schurComp}(D_k, \epsilon)$.

\subsubsection{Modified Cholesky in Diagonal Tiles}

A different approach to introduce ``minimal'' perturbations to insure positive definiteness consists of modifying an offending diagonal tile $A(k,k)$ when it loses its definiteness
by adding a small-normed symmetric perturbation $E$ \cite{cheng_higham_modchol_1998}.  The solution is a modified Cholesky algorithm that first produces a symmetric indefinite tile factorization
\begin{equation*}
P A(k,k) P^T = L D L^T.
\end{equation*}
$D$ is then modified with perturbations $F$ to make the sum $D + F$ positive 
definite, providing the factorization of an augmented positive definite matrix $A + E$
\begin{equation*}
A(k,k) + E = P^T L (D + F) L^T P.
\end{equation*}
When this strategy succeeds and the norm of $E$ is sufficiently small, the diagonal block
 can be replaced by the modified one and the regular Cholesky can be carried out.
This is a small modification to Line \ref{alg:piv_tlr_chol:ldl_update} 
of Algorithm \ref{alg:piv_tlr_chol}, replacing the regular dense Cholesky with the modified one in Algorithm 
\ref{alg:mod_chol}.
\begin{algorithm}
\caption{Modified Cholesky}
\label{alg:mod_chol}
\begin{algorithmic}[1]
\Procedure{ModChol}{$A$}
	\Input
	\Desc{$A$}{Dense Matrix}
	\EndInput	
	\Output 
	\Desc{$L$}{Cholesky factor of $A+E$. $E = 0$ if $A$ is positive definite.}
	\EndOutput
	
	\State $[L, i] = $ chol $(A)$	\Comment{$i$ is \texttt{fail} if the factorization fails}
	\If{$i = \texttt{fail}$}
		\State $[L, D, P] = $ ldl $(A)$
		\State $F = $ modifySPD $ (D) $ 		\Comment{Modify $D$ to make $D+F$ positive definite}
		\State $\tilde{A} = P^T L (D + F) L^T P$	
		\State $L = $ chol $(\tilde{A})$
	\EndIf
\EndProcedure
\end{algorithmic}
\end{algorithm}


\subsection{Inter-tile Pivoting}

In the context of TLR, two types of pivoting are possible. The first is \emph{inter-tile}, where the tiles are swapped but their contents remain untouched, while the other is \emph{intra-tile} pivoting, which is the standard pivoting associated with scalar factorizations. For the factorization of its diagonal dense blocks, our algorithm relies on LAPACK routines. We assume that intra-tile pivoting is handled at that level. In this section we briefly describe our considerations for inter-tile pivoting. In the next section, we show sample results for the effect of inter-tile pivoting on performance, including indirect effects due to rank changes in tiles as a result of pivoting. 


For semidefinite and definite cases, we only consider a symmetric pivoting with the pivot tile chosen 
from the diagonals. This symmetric tile pivoting is fairly simple to accomplish, can have 
positive effects on the overall stability of the TLR Cholesky, and is an important step towards a more 
robust algorithm. In the scalar pivoting case, where it is possible to swap individual rows and 
columns, the diagonal entry $p$ with the largest value is selected at step $k$ of the algorithm and the 
$p$ and $k$ rows and columns are swapped. This ensures that the update to the trailing sub-matrix has a small norm. 
We use a similar heuristic for TLR matrices, 
selecting the diagonal tile that has the largest norm. The 2-norm can be approximated using power iterations, 
but this can be quite costly. When possible, an alternative is to use the Frobenius norm, which can be evaluated 
 more efficiently, especially on GPUs where power iterations for many small tiles can perform quite 
poorly. 

Once the tile pivot is selected, it is straightforward to simply swap pointers around and proceed with the 
updates as before. However, the dense update changes slightly in the pivoted algorithm, since we need to select 
the pivot based on the updated diagonal blocks, which in a left looking scheme are only updated once. 
This entails having a temporary set of diagonal tile updates that contain the sum of all low rank updates 
$D_i$ to the diagonal blocks $A(i, i)$, allowing the selection of the pivot $p = \texttt{argMax}_{i=k:n}(||A(i,i) 
- D_i||)$ at step $k$. While this does increase the workspace memory costs of the algorithm, it has the 
positive effect of increasing parallelism, as all $D_i$ can be updated in parallel. Algorithm \ref{alg:piv_tlr_chol} 
shows the pivoted tile low rank Cholesky, which is almost identical to Algorithm \ref{alg:tlr_chol} with 
the exception of the updates to the diagonal and pivot selection (lines \ref{alg:piv_tlr_chol:update_1}-\ref{alg:piv_tlr_chol:update_2}).

\begin{algorithm}
\caption{Pivoted Left Looking TLR Cholesky}
\label{alg:piv_tlr_chol}
\begin{algorithmic}[1]
\Procedure{PivTLRChol}{$A, bs, \epsilon, ws$}
	\Input
	\Desc{$A$}{TLR Matrix}
	\Desc{$bs$}{Block samples for ARA}
	\Desc{$ws$}{Workspace}
	\EndInput
	
	\Output 
	\Desc{$A$}{Lower triangular matrix $L$ overwriting $A$}
	\Desc{$p$}{Permutation array}
	\EndOutput
	
	\State $nb = $ blocks$(A)$
	\For{$k = 1 \rightarrow nb$}
		\State $D = $ denseUpdates $(A, k, ws)$   \Comment{Update the sums of low rank updates} 
		\label{alg:piv_tlr_chol:update_1}
		\State $p = $ selectPivot $(A, D, k, p)$  \Comment{Select a pivot tile}
		\State $A = $ pivotRowColumn $(A, k, p)$  \Comment{Swap rows and columns}
		\State $A(k, k) = A(k, k) - D_k$ 		  \Comment{Dense update}
		\label{alg:piv_tlr_chol:update_2}
		\State $A(k, k) = $ chol $(A(k, k))$	  \Comment{Dense Cholesky}
		\label{alg:piv_tlr_chol:ldl_update}
		\State $[Q, B] = $ cholARAUpdate $(A, bs, k, \epsilon, ws)$  \Comment{Approximate updated column tiles}

		\State $B = $ batchTrsm $(A(k, k), B, nb - k - 1)$ \Comment{Triangular solve}
		\State $A = $ updateTiles $(A, k, Q, B)$ \Comment{Replace old tiles with the updated ones}
	\EndFor
\EndProcedure
\end{algorithmic}
\end{algorithm}

\subsection{LDL$^T$ Factorization}
For the indefinite case, a small set of simple modifications to the Cholesky TLR algorithm can be used to compute the (unpivoted) $LDL^T$ TLR factorization. First, the factorization of the block diagonals should be carried out as dense $LDL^T$ factorization instead of dense Cholesky: $A(k,k)= L(k,k) D(k,k) L(k,k)^T$. The triangular solves below the diagonal must then be updated to incorporate the diagonal matrix $D(k,k)$, transforming the right low rank factors $B = D(k,k)^{-1} B$ in an inexpensive diagonal scaling step. Finally, the dense updates and sampling need to be modified as a sequence of five matrix-matrix products instead of four, transforming Equation \ref{eq:sample_lru} into:
\begin{equation}
\label{eq:sample_ldl_lru}
Y_j = U(i, j) \left(V(i,j)^T \left( D(j,j) \left( V(k, j) \left( U(k,j)^T \Omega_j \right) \right) \right) \right).
\end{equation}
The remaining ARA steps remain the same, producing Algorithm \ref{alg:tlr_ldl} as a modification to Algorithm Algorithm \ref{alg:tlr_chol}. 

Pivoting for the indefinite case is tricky for TLR matrices. The inter-tile diagonal pivoting of the definite case, even coupled with pivoting within tiles, is not in principle sufficient, unless the tiles have been organized in a way that provides guarantees a priori. Further work is needed to analyze pivoting requirements for the indefinite case. However, interesting alternatives to pivoting are possible \cite{becker11}. A symmetric randomization of the matrix with recursive butterfly matrices appear to provide the stability needed for indefinite factorization to succeed without pivoting. The random butterfly transformation is ideal for GPU implementation and we hope to explore this direction in future work.

\begin{algorithm}
\caption{Left Looking TLR $LDL^T$}
\label{alg:tlr_ldl}
\begin{algorithmic}[1]
\Procedure{TLRLDL}{$A, bs, \epsilon, ws$}
	\Input
	\Desc{$A$}{TLR Matrix}
	\Desc{$bs$}{Block samples for ARA}
	\Desc{$ws$}{Workspace}
	\EndInput
	
	\Output 
	\Desc{$A$}{Lower triangular matrix $L$ overwriting $A$}
	\Desc{$D$}{Diagonal matrix $D$ such that $A = LDL^T$}
	\EndOutput
	
	\State $nb = $ blocks$(A)$
	\For{$k = 1 \rightarrow nb$}
		\State $A(k, k) = $ updateDenseLDL $(A, D, k, ws)$ \Comment{Dense updates of diagonal block}
		\State $[A(k, k), D(k,k)] = $ ldl $(A(k, k))$	     \Comment{Dense $LDL^T$}
		\State $[Q, B] = $ ldlARAUpdate $(A, D, bs, k, \epsilon, ws)$  \Comment{Approximate updated column tiles}

		\State $B = $ batchTrsm $(A(k, k), B, nb - k - 1)$ \Comment{Triangular solve}
		\State $B = $ batchIDMult $(D(k,k), B)$ \Comment{Inverse diagonal multiply}
		
		\State $A = $ updateTiles $(A, k, Q, B)$ \Comment{Replace old tiles with the updated ones}
	\EndFor
\EndProcedure
\end{algorithmic}
\end{algorithm}

\section{Performance Results}

In this section we discuss the performance of the proposed algorithm using covariance matrices arising from spatial 
Gaussian processes in two and three dimensions and an isotropic exponential kernel with correlation lengths 
of $0.1$ and $0.2$ respectively \cite{ambi16}. We also assess the quality of the computed low-rank factorization
as a preconditioner for a discretization of an integral equation formulation of 
space-fractional diffusion equations in three dimensions \cite{BOUKARAM2020113191}. 
The test system has a dual socket 20-core Intel(R) Xeon(R) Gold 6148 CPU with 384 GB of system memory 
and an NVIDIA V100 GPU with 16GB of memory. All timing results are from averages of 10 runs using double precision. 
On the CPU, we use the MKL 2020 library with OpenMP for the parallelization of batched routines, using 20 threads 
and the dynamic scheduler.
On the GPU, we use the non-uniform batched matrix-matrix kernels from MAGMA 2.5.4 \cite{magma} 
together with H2Opus for the remaining batch kernels \cite{h2opus}.
GPU performance is still partially limited by sub-optimal occupancy, 
since the needed MAGMA kernels are not yet fully asynchronous due to
memory allocations and deallocations and device to host synchronous copies,
further preventing overlapping of the low rank and dense updates.
On the CPU, the batch ecosystem with variable sizes is not yet as mature and performant as in the GPU case;
speed ups will be readily available from future performance improvements offered by specialized batched APIs \cite{heinecke2016libxsmm}.

The tunable parameters for the algorithm, namely the 
number of parallel buffers used for sampling and dense updates, were set to $\frac{3}{2} b$ and $20$ 
respectively, leading to a workspace size asymptotically lower than the memory required by 
the matrix itself, and to an acceptable compromise between parallelism and memory consumption in the specific tests considered here.
Guided  by the observation that the ranks in the 3D problems were typically larger than those of the 2D problems, 
the ARA sampling block size was set to $16$ for 2D and $32$ for 3D problems. Unless otherwise stated, the tile sizes
used for the covariance matrices in two and three dimensions are $1024$ and $512$ respectively, and the data
points are uniformly distributed in a grid. 
All results were verified by estimating the 2-norm of the difference $||A-LL^T||$ using the power iteration method. 

\begin{figure}
	\begin{subfigure}[t]{0.49\textwidth}
	\centering
	\includegraphics[width=\textwidth]{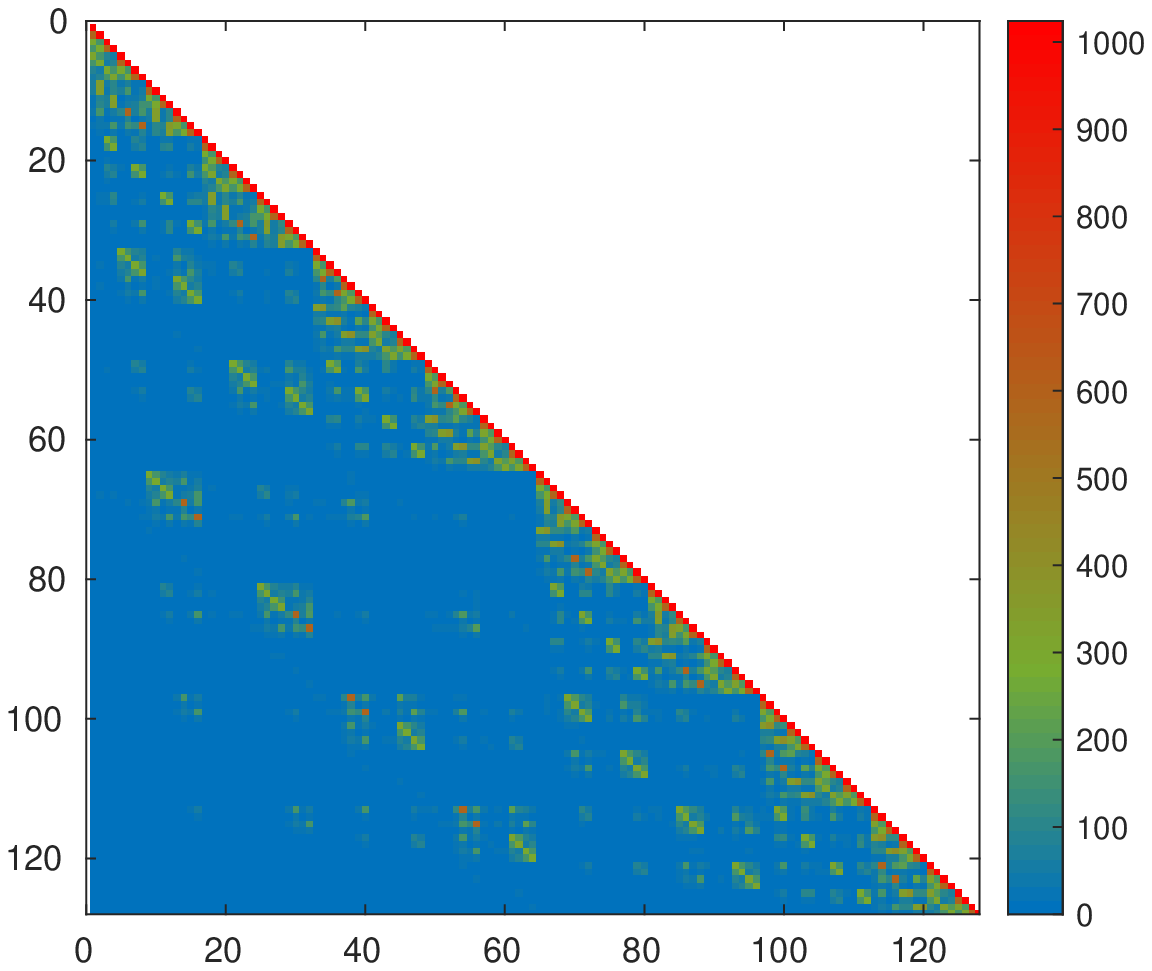}
	\caption{Fractional diffusion.}
	\label{fig:heatmap_L_fd}
	\end{subfigure}
	\hfill	
	\begin{subfigure}[t]{0.49\textwidth}
	\centering
	\includegraphics[width=\textwidth]{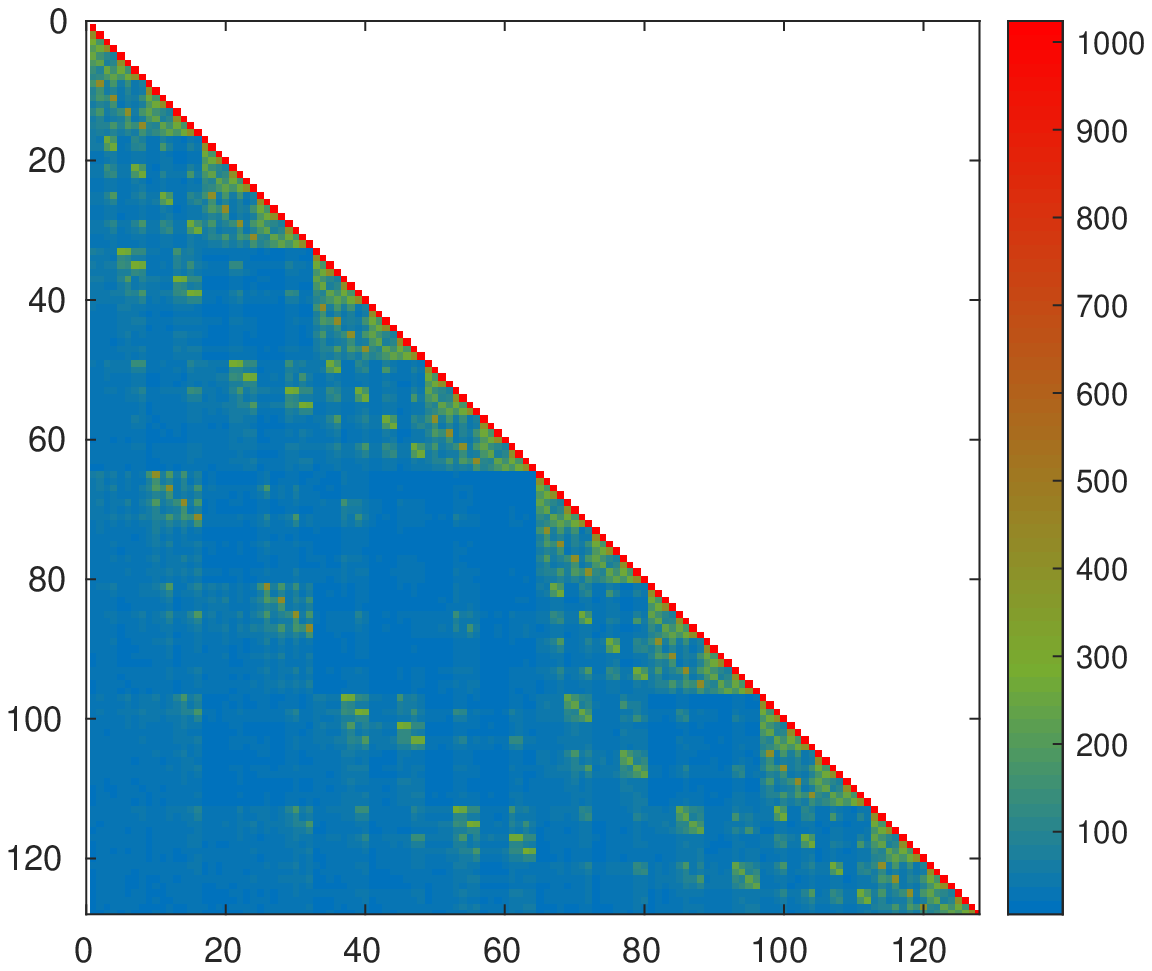}
	\caption{Covariance.}
	\label{fig:heatmap_L_cov}
	\end{subfigure}
	\caption{Heatmaps of the ranks of the tiles for the Cholesky factors of fractional diffusion and covariance 
matrices of size $N=2^{17}$ and a tile size of $1024$ using an absolute compression threshold of $10^{-6}$.}

	\label{fig:heatmap_L}
\end{figure}

The ordering of the geometric data was determined by partitioning the $N$ geometric points in each problem using 
a KD-tree where plane splits aimed to partition points into clusters that are as close to the chosen 
tile size as possible. The points within each cluster were sorted by projecting along the largest dimension 
of its bounding box and then split into a left cluster whose size is half the closest power of two of 
the full cluster multiplied by the tile size and a right cluster containing the remaining points. This 
produces a cluster tree whose leaves are all the same size with the possible exception of the right most 
leaf, allowing the construction of the tile low rank matrix with just the final block row and column 
requiring padding. Other clustering techniques based on space-filling curves could be used to determine 
the ordering which reduces the rank the most, but we leave that determination as future work. Figures 
\ref{fig:heatmap_L_fd} and \ref{fig:heatmap_L_cov} show heatmaps of the resulting tile low rank Cholesky 
factors of three dimensional fractional diffusion and covariance matrices respectively. Each matrix is 
of size $N=2^{17}$ with a tile size of $1024$ and constructed to an absolute compression threshold of 
$10^{-6}$, which, when scaled by their 2-norms, gives relative errors of about $10^{-5}$ and $10^{-9}$ 
respectively.

The remainder of this section is organized as follows. Section \ref{sec:res_cov} discusses the performance of the algorithm for various covariance matrices.
Section \ref{sec:res_fd} discusses the performance of the factorization and its effectiveness as a preconditioner for solving fractional diffusion problems.
Section \ref{sec:res_extensions} shows the performance costs and implications for using pivoted Cholesky and the $LDL^T$ factorization over ``vanilla'' TLR Cholesky.

\subsection{Performance on Covariance Matrices}  
\label{sec:res_cov}

\begin{figure}
	\begin{subfigure}[t]{0.45\textwidth}
	\centering
	\includegraphics[width=\textwidth]{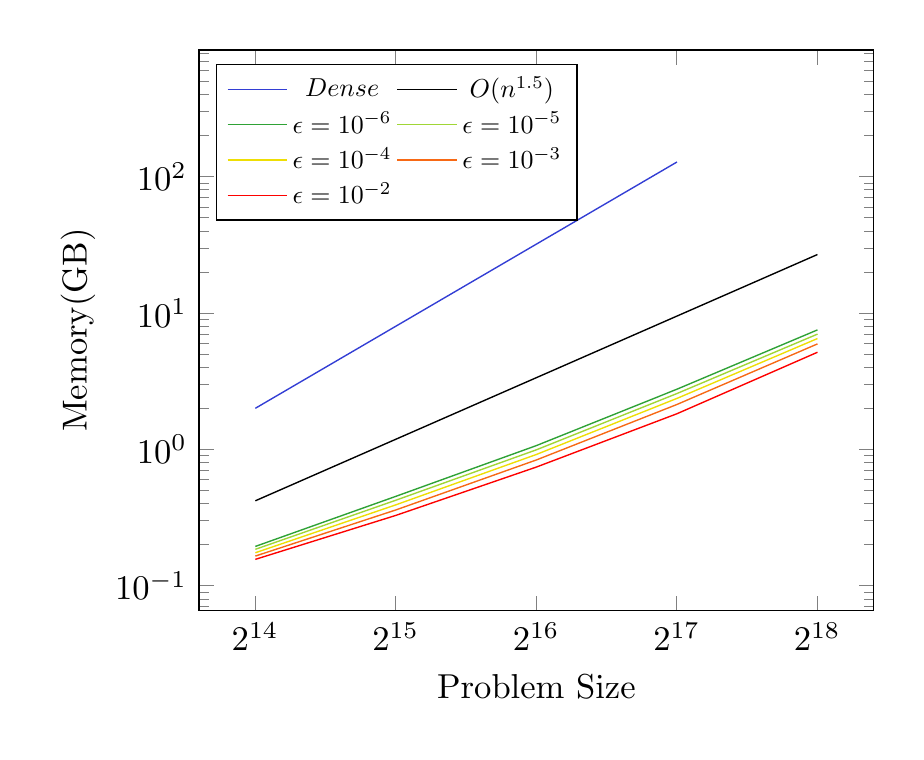}
	\caption{2D covariance matrix.}
	\label{fig:mem2d}
	\end{subfigure}
	\hfill	
	\begin{subfigure}[t]{0.45\textwidth}
	\centering
	\includegraphics[width=\textwidth]{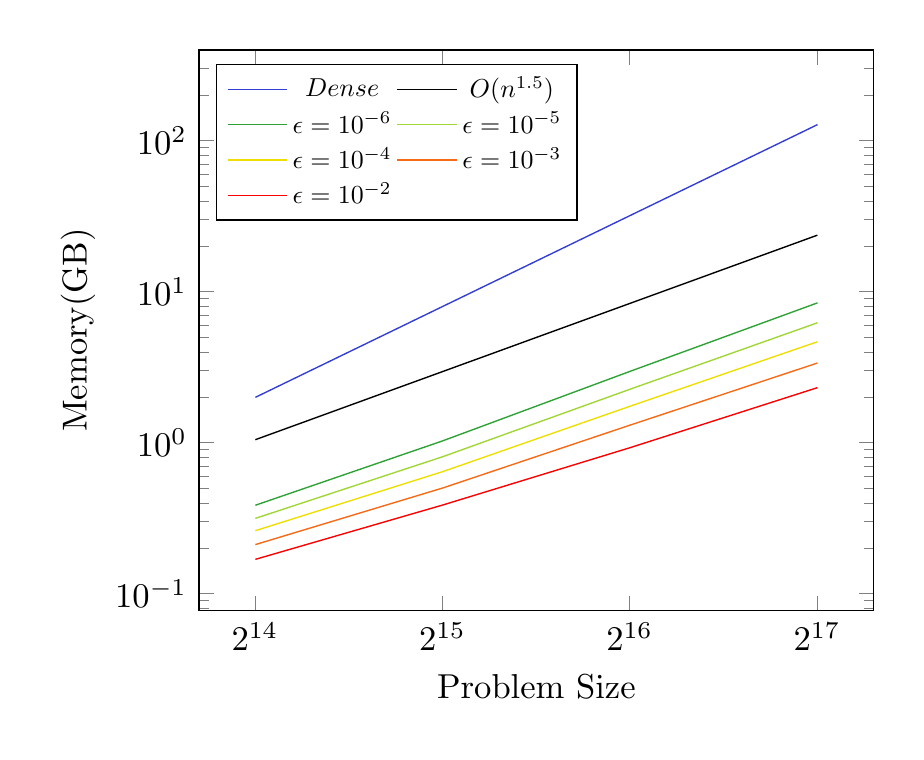}
	\caption{3D covariance matrix.} 
	\label{fig:mem3d}
	\end{subfigure}
	\caption{Memory growth for 2D and 3D covariance matrices. Note that the lower ranks in 2D allow larger problems to fit in 
the 16GB of GPU memory.}
	\label{fig:mem_vs_size}
\end{figure}

One of the main benefits of using tile low rank matrices is the asymptotically lower memory consumption 
when compared to the full dense representation. Figures \ref{fig:mem2d} and \ref{fig:mem3d} show the 
growth of total memory consumption of the TLR representation as the problem size increases
for various compression thresholds $\epsilon$.
The observed asymptotic memory growth $O(n^{1.5})$ of the tile low rank format is in agreement with the estimates \cite{amestoy17}
and it contrasts favorably with the $O(n^2)$  growth of the dense representation, especially on memory starved architectures like GPUs.
Even with ample main system memory, the memory requirements for the full dense representation prevent working with larger 
problems sizes, whereas the relatively meager 16GB of the V100 GPU can fit even larger problems if the 
accuracy requirements or the compressed ranks aren't too high. This can be clearly seen in Figure \ref{fig:mem_vs_size}, where the memory required from the overall higher ranks with $N=2^{18}$ exceeds the available memory; more favorable rank
distributions in two dimensions instead allow us to represent matrices of this size.

\begin{figure}
	\begin{subfigure}[t]{0.45\textwidth}
	\centering
	\includegraphics[width=\textwidth]{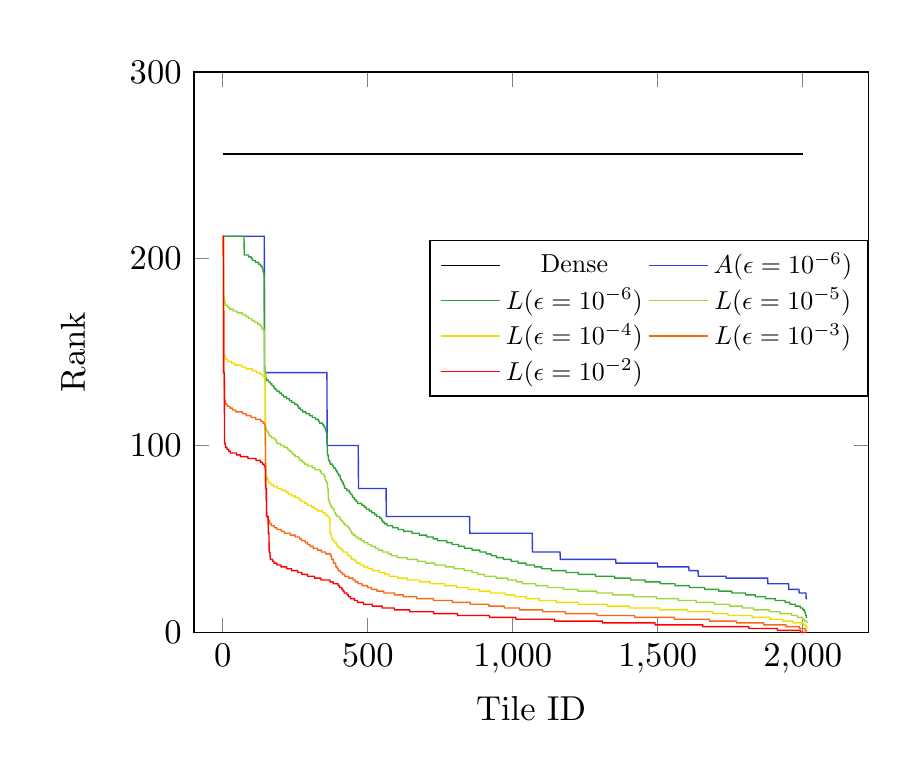}
	\caption{Rank distribution for a 3D regular grid.}
	\label{fig:rank_grid}
	\end{subfigure}
	\hfill	
	\begin{subfigure}[t]{0.45\textwidth}
	\centering
	\includegraphics[width=\textwidth]{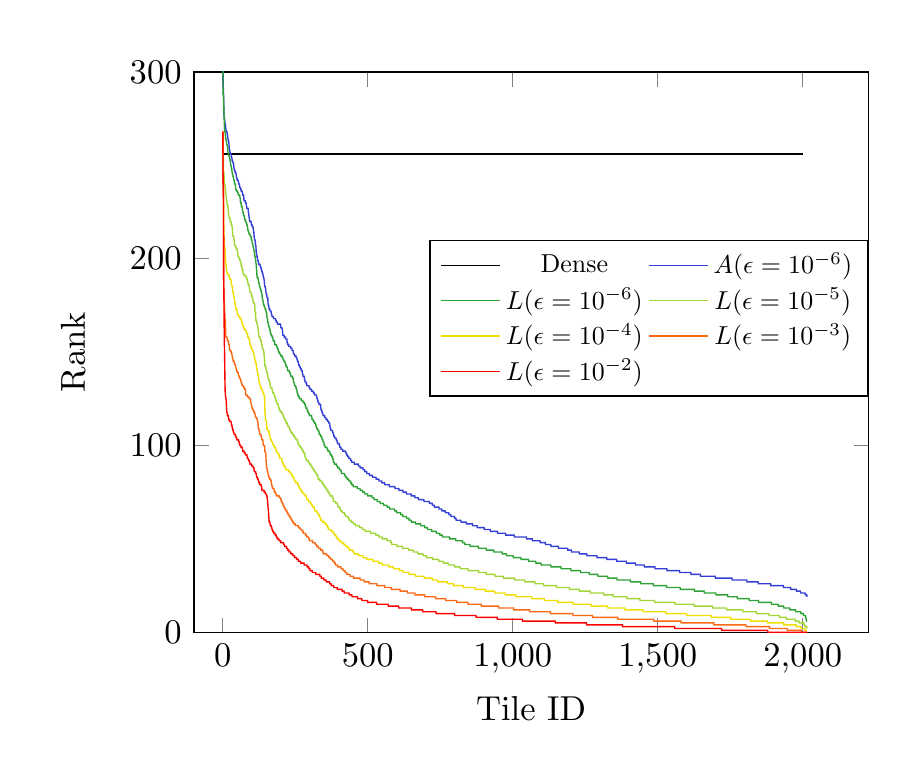}
	\caption{Rank distribution for random points in a 3D ball.}
	\label{fig:rank_sphere}
	\end{subfigure}
	\caption{Rank distribution for two problems for a 3D $N=2^{15}$ covariance matrix with a tile size of 
$512$ revealing data sparsity and possible wasted memory due to ranks greater than half the tile size.}

	\label{fig:rank_dist}
\end{figure}

To better understand the source of the memory savings of the factorizations with different compression thresholds $\epsilon$,
we visualize the rank distribution of the tiles of the TLR matrix and its Cholesky factors for a three dimensional $N=2^{15}$ covariance 
matrix for a uniform point distribution in Figure \ref{fig:rank_grid}, and for a
random distribution of points in a 3D ball in Figure \ref{fig:rank_sphere}.
In addition to the distribution of the ranks of the tiles, these figures 
encapsulate many aspects of the TLR matrix, the most obvious of which is the data sparsity of the representation.
The area under each curve can be interpreted as a proxy for the level of compression of the TLR matrix 
when compared to the dense representation (black line).
Storing blocks with ranks greater 
than half the tile size in low-rank format introduces some overhead, in that the memory required by
the low-rank format exceeds that of the full dense representation.
The area above the dense black line and below the rank distribution curves thus represent the amount of
this memory overhead required by not allowing dense blocks to appear on the off-diagonal tiles of the TLR representation
in the spirit of maintaining a simpler code.
The TLR representation on the regular grid does not incur any overhead and it features many tiles having the same rank,
whereas the extra memory for the random distribution of points on a sphere is negligible with respect to the 
overall TLR memory consumption.

\renewcommand{\arraystretch}{1.3}
\setlength\tabcolsep{3pt}

\begin{table}[]
\begin{tabular}{c|c|c|c|c|c|c|c|c|}
\cline{2-9}
\multicolumn{1}{l|}{}           & \multicolumn{4}{c|}{$N=2^{15}$}                 & \multicolumn{4}{c|}{$N=2^{16}$} 
                \\ \hline
\multicolumn{1}{|c|}{Tile Size} & Total  & Dense  & Low Rank & Cholesky & Total  & Dense  & Low Rank 
& Cholesky \\ \hline
\multicolumn{1}{|c|}{128}       & 1.42 & 0.03 & 1.39   & 12.49  & 4.16 & 0.06 & 4.10   
& 56.6  \\ \hline
\multicolumn{1}{|c|}{256}       & 1.18 & 0.06 & 1.12   & 10.24  & 3.42 & 0.12 & 3.30   
& 36.2  \\ \hline
\multicolumn{1}{|c|}{512}       & 1.02 & 0.12 & 0.90   &  6.98  & 2.97 & 0.25 & 2.72   
& 26.8  \\ \hline
\multicolumn{1}{|c|}{1024}      & 1.13 & 0.25 & 0.88   &  8.86  & 2.72 & 0.50 & 2.22   
& 22.3  \\ \hline
\multicolumn{1}{|c|}{2048}      & 1.35 & 0.50 & 0.85   & 14.22  & 3.25 & 1.00 & 2.25   
& 32.3  \\ \hline
\end{tabular}
\caption{The effect of varying the tile size on memory (in GB) and runtime (in seconds) for two 3D covariance matrices on the CPU. Factorization was done with a threshold $\epsilon = 10^{-6}$.}
\label{tab:vary_tile_size}
\end{table}

Tile size has a large impact on memory consumption and factorization times and depends on the rank
distribution and size of the matrix being approximated. If there are many instances where a tile
with large rank is adjacent to tiles with significantly smaller ranks, then merging those tiles
by increasing the tile size will lead to greater overall memory consumption. 
On the other hand, and in order to maintain the asymptotic performance estimates of the
TLR representation \cite{amestoy17}, the tile size should also
increase with the problem size as $O(N^{0.5})$. Size selection is also affected 
by the cache size where the tiles reside in the memory hierarchy. 
 Table \ref{tab:vary_tile_size} shows the effect of increasing the tile size
on the overall memory consumption and factorization times of two sample matrices. It is clear that the
tile size for the smaller problem should be around $512$ and for the larger problem it should be increased
to about $1024$.
It is generally difficult to determine the ideal tile size for a given problem without a priori information
on the rank distribution of the matrix, so for the results listed in this section we determined reasonable
tile sizes empirically.

\begin{figure}
	\begin{subfigure}[t]{0.45\textwidth}
	\centering
	\includegraphics[width=\textwidth]{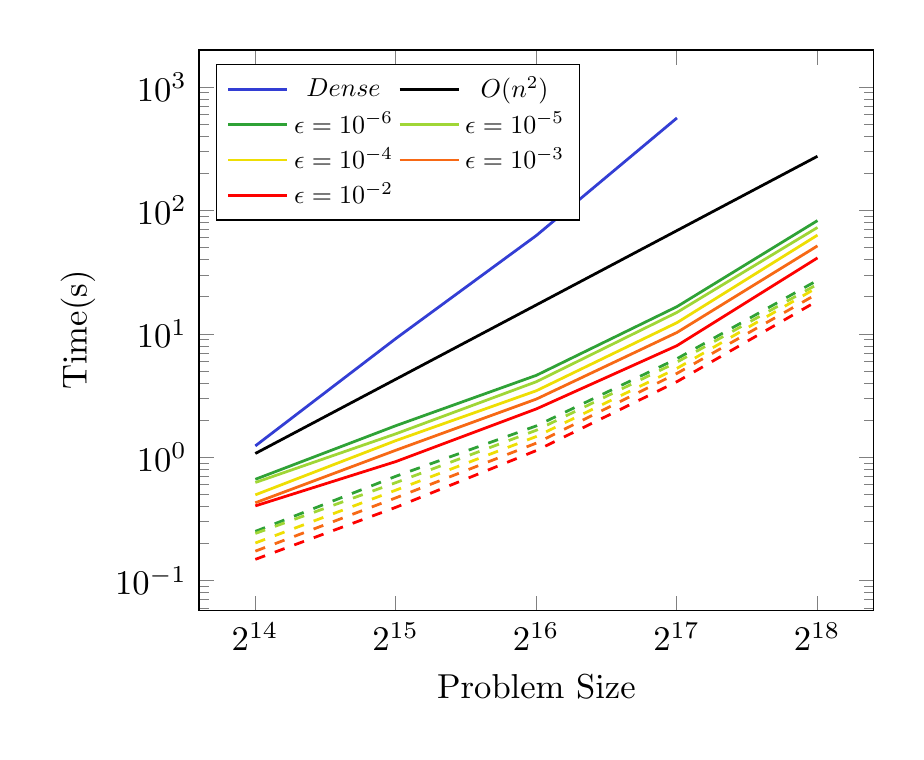}
	\caption{Time for two dimensional covariance matrices with a tile size of $1024$}
	\label{fig:time_2d}
	\end{subfigure}
	\hfill	
	\begin{subfigure}[t]{0.45\textwidth}
	\centering
	\includegraphics[width=\textwidth]{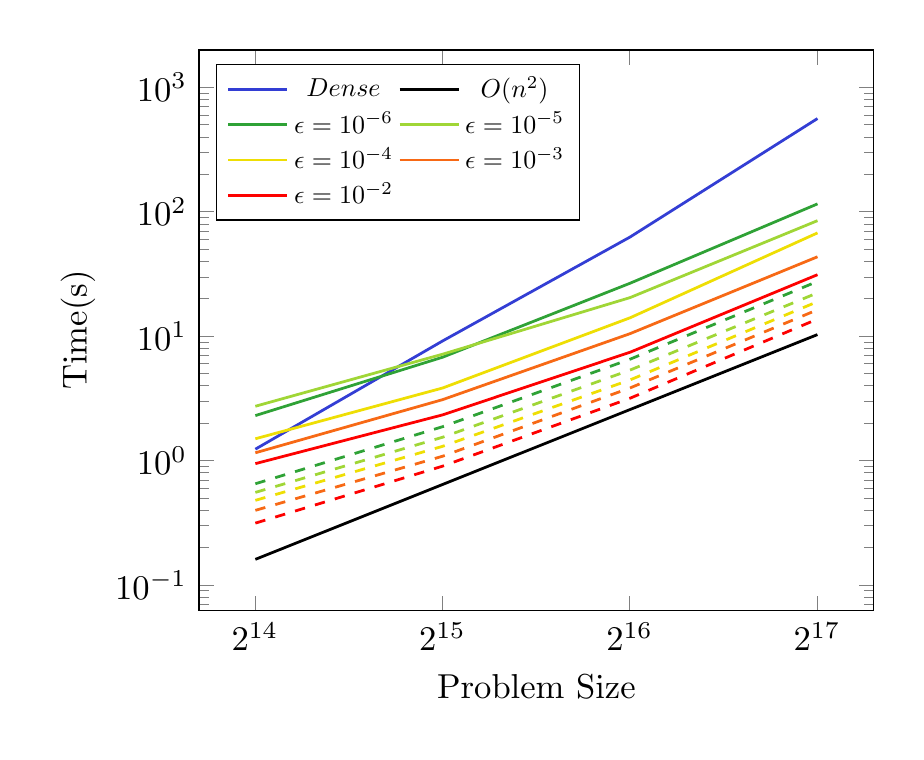}
	\caption{Time for three dimensional covariance matrices with a tile size of $512$.}
	\label{fig:time_3d}
	\end{subfigure}
	\caption{TLR Cholesky factorization times for 2D and 3D covariance problems. Solid lines 
show CPU times and dashed lines show GPU times.}
	\label{fig:time_vs_size}
\end{figure}

Figures \ref{fig:time_2d} and \ref{fig:time_3d} plot the time taken for the TLR Cholesky for 
various problem sizes and compression thresholds on the GPU (dashed lines) and on the CPU (continuous lines).
The timings associated with the dense representation have been obtained using the threaded MKL
library and they are provided for comparison.
When compared against the dense factorization, the significantly lower ranks 
of the two dimensional problems lead to far greater time savings than those obtained in the three 
dimensional case.
However, the superior asymptotic operational complexity of the TLR Cholesky on the CPU leads to an almost 5x 
improvement in factorization time for the 3D problem and a 32x improvement for the 2D problem with the tighest accuracy $\epsilon = 10^{-6}$.
For lower accuracies, the gains are even larger at 
about 17x for the 3D and 69x for the 2D problems for an absolute accuracy of $\epsilon = 10^{-2}$.
The GPU implementation shows a 2--5x improvement over the CPU TLR times; this is once more largely 
dependent on the rank distribution of the TLR matrix, where low ranks result in lower computational intensity 
and overall less efficiency from the non-uniform batched GPU matrix-matrix multiplication kernels.

\begin{figure}
	\begin{subfigure}[t]{0.51\textwidth}
	\centering
	\includegraphics[width=\textwidth]{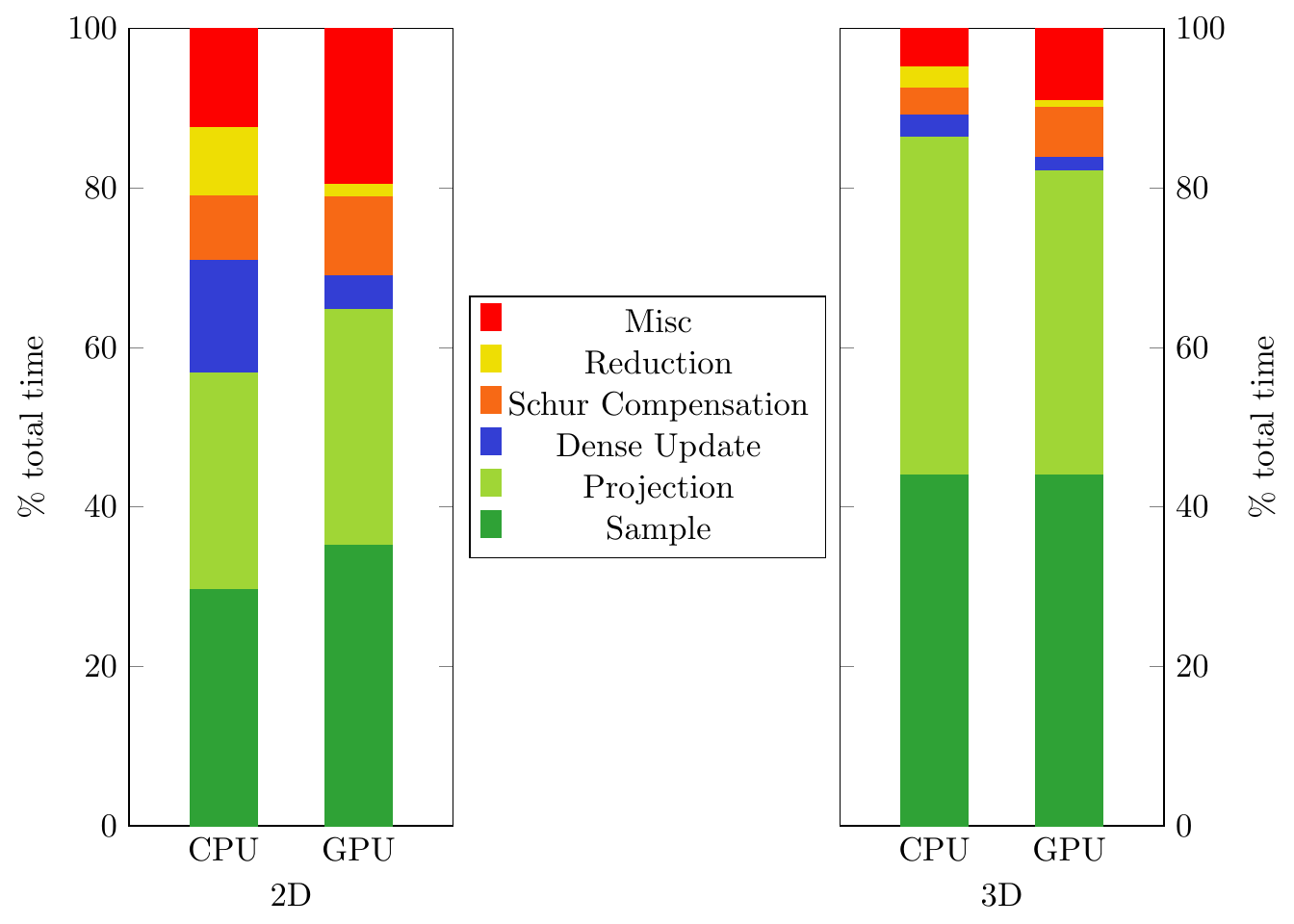}
	\caption{TLR Cholesky profile for 2D and 3D covariance matrices of size $N=2^{17}$ and compression threshold 
$\epsilon=10^{-6}$ showing the dominance of matrix-matrix multiplication operations on the CPU and GPU.}

	\label{fig:tlr_chol_profile}
	\end{subfigure}
	\hfill	
	\begin{subfigure}[t]{0.47\textwidth}
	\centering
	\includegraphics[width=\textwidth]{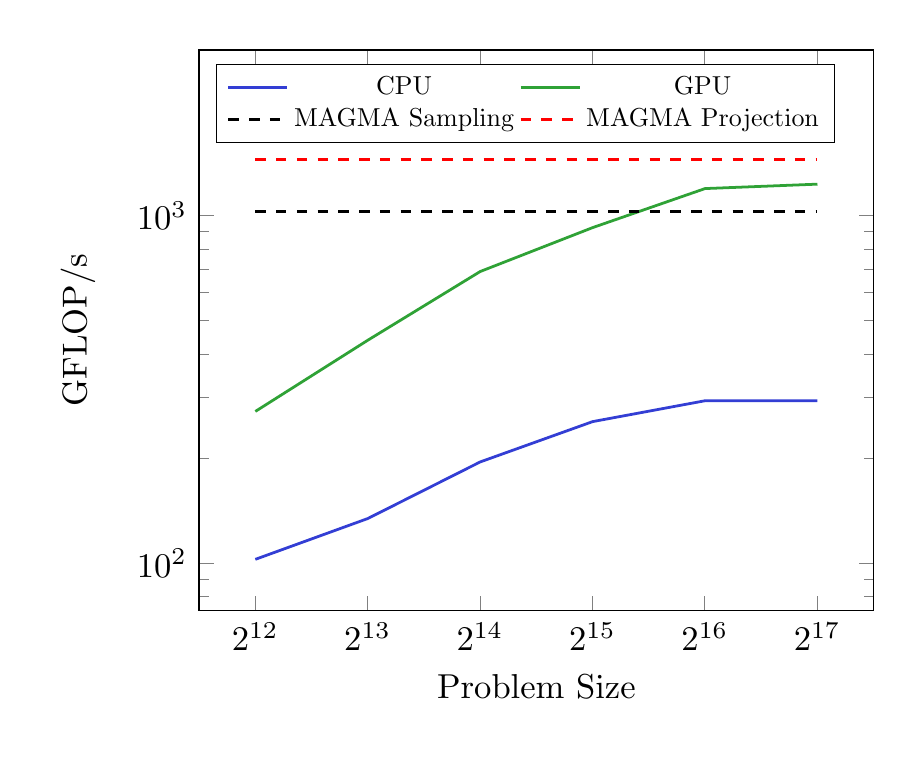}
	\caption{Performance of the TLR factorization for various 3D covariance matrices compressed to $\epsilon=10^{-6}$. 
  Dashed lines represent the 
 limits of the batched matrix-matrix products in sampling and projection for representative ranks.}
	\label{fig:flops}
	\end{subfigure}
	\caption{Performance profiles for matrices of size $N=2^{17}$ and overall performance of the factorization 
on the CPU and GPU.}
	\label{fig:flops_and_profile_cov}
\end{figure}

Figure \ref{fig:tlr_chol_profile} provides a breakdown of the timings associated with the different phases of the factorizations for $N=2^{17}$ and using a compression threshold $\epsilon=10^{-6}$.
The miscellaneous part includes
the dense diagonal block factorizations, triangular solves, random number generation, memory reallocations, 
basis orthogonalization and operation marshalling. Many of these operations are less efficient on the 
GPU when compared to the matrix-matrix multiplications and therefore take up a larger portion of the 
overall GPU runtime. 
The smaller ranks featured in the two dimensional test further increase the relative costs of these
operations up to a 20\% on the GPU, mainly due to the dense block factorizations with fixed costs given the tile size.
With the exception of the reduction phases, 
all the other kernels are based on small matrix-matrix products of different sizes.
In total, these high-efficiency kernels represent about 80-90\% of the total factorization times either on the CPU or the GPU cases.

Figure \ref{fig:flops} shows the 
overall performance of the factorization on the CPU and GPU in terms of floating point operations per seconds for the three-dimensional case. 
In order to assess the quality of the implementantion and obtain a rough estimate of the expected GPU
performance for the most time consuming operations of sampling and projection,
we benchmarked the MAGMA non-uniform batched matrix-matrix
product kernels and average results with the $m \times k$ times $k \times n$ ($AB$) kernel and the $(k \times m)^T$ times $k \times n$ ($A^TB$) kernel, with $m, n$ and $k$ chosen close to those used 
by the algorithm in each phase. For sampling, given that the ARA block size is $32$, we set $m=512$, $n=32$ and $k$ uniformly distributed between $16$ and $48$, since
the majority of the ranks are in this range. On the other hand, since the projection phase uses 
the rank detected by the ARA, we benchmark this phase by considering a uniformly distributed $n$ in the same range as well. In both cases, the batch size is set to $500$ as a rough 
upper bound on the number of parallel buffers used in the workspace. The figure shows that GPU performance 
asymptotically lands neatly between the two rough estimates. 

\subsection{Performance on Fractional Diffusion}  
\label{sec:res_fd}
\begin{figure}
	\centering
	\includegraphics[width=0.6\textwidth]{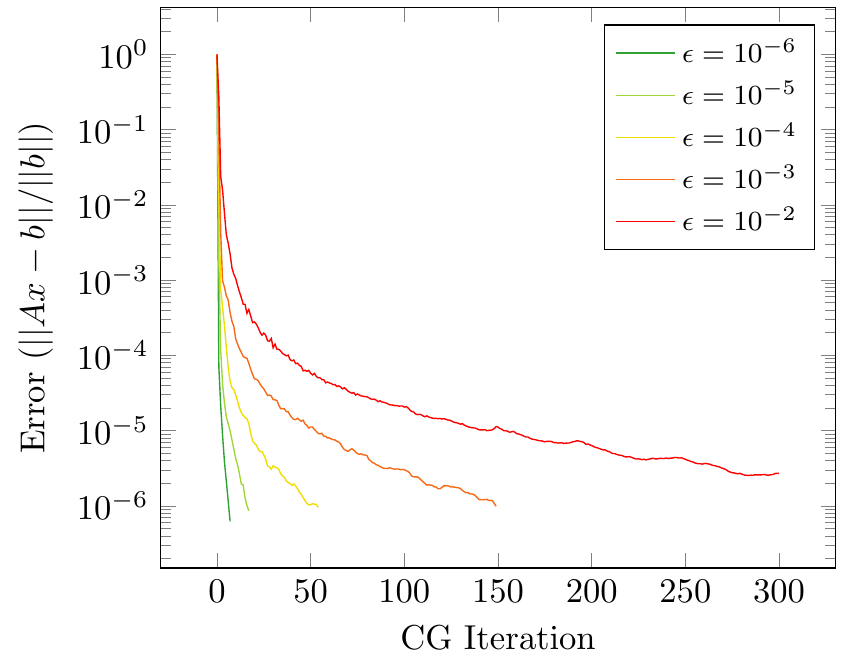}
	\caption{Preconditioned CG convergence for various compression thresholds $\epsilon$ using the factorization 
of $A+\epsilon I$ as the preconditioner.}
	\label{fig:tlr_chol_cg}
\end{figure}

\begin{figure}
	\begin{subfigure}[t]{0.4\textwidth}
	\centering
	\includegraphics[width=\textwidth]{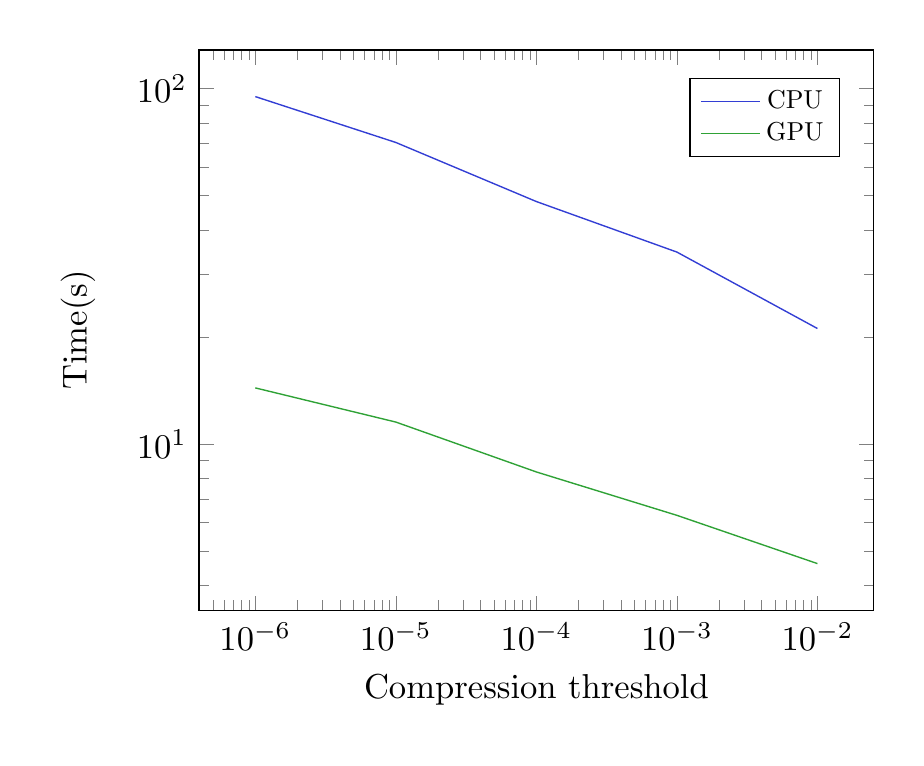}
	\caption{Time taken to generate the TLR factorization (preconditioner) for various compression thresholds.}
	\label{fig:time_vs_eps_3D_fd}
	\end{subfigure}
	\hfill	
	\begin{subfigure}[t]{0.58\textwidth}
	\centering
	\includegraphics[width=\textwidth]{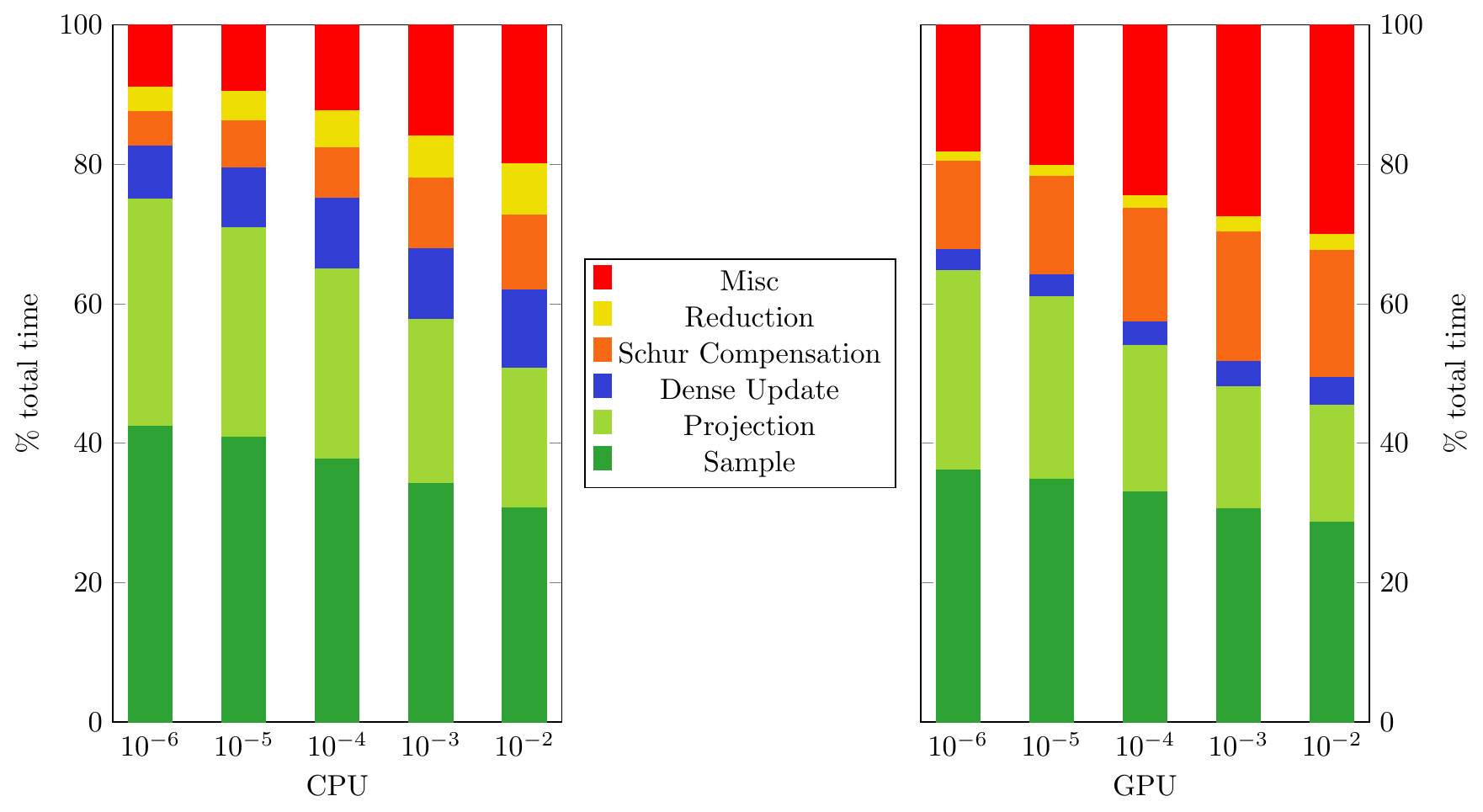}
	\caption{Percentage of time spent in each phase for various compression thresholds.}
	\label{fig:profile_fd_eps}
	\end{subfigure}
	\caption{Performance results for constructing a preconditioner for a 3D fractional diffusion matrix 
of size $N=2^{17}$.}
	\label{fig:fd_eps}
\end{figure}

When the matrix is ill-conditioned, as is the case with the fractional diffusion matrix which has a 
condition number of about $10^7$ for $N=2^{17}$, it is often 
more effective to generate a low accuracy factorization to use as a preconditioner for iterative solvers 
like the Conjugate Gradient (CG) method. Figure \ref{fig:tlr_chol_cg} shows the number of preconditioned 
CG iterations required to converge to an error in the solution $||A x - b||/||b|| = 10^{-6}$ for a 3D 
fractional diffusion matrix of size $N=2^{17}$, where the TLR Cholesky factorization of the matrix $A 
+ \epsilon I$ was compressed at various thresholds $\epsilon$ and used as a preconditioner. The matrix $A$ 
is quite ill-conditioned. The addition of the identity scaled by the compression threshold maintains
positive definiteness. Since we are adding a term on the same order of the compression threshold, 
the resulting factorization error remains close to the threshold. The lowest accuracy factorization fails 
to provide a preconditioner that 
converges within the maximum number of iterations ($300$), while higher accuracy 
lead to higher convergence rates. Figure 
\ref{fig:time_vs_eps_3D_fd} shows the factorization time for the TLR matrix on the CPU and GPU for various 
compression thresholds, while Figure \ref{fig:profile_fd_eps} breaks down the percentage of time spent 
in each phase of the computation. The amount of time spent in the high efficiency kernels decreases as 
the rank decreases with the accuracy, but even at the lowest accuracy they comprise almost 70\% of the 
computation time. The TLR matrix vector products and triangular solves complete quickly relative 
to the factorization time, taking 0.177 and 0.385 seconds respectively on the CPU and 0.068 and 0.15 seconds on the GPU.

\begin{figure}
	\begin{subfigure}[t]{0.49\textwidth}
	\centering
	\includegraphics[width=\textwidth]{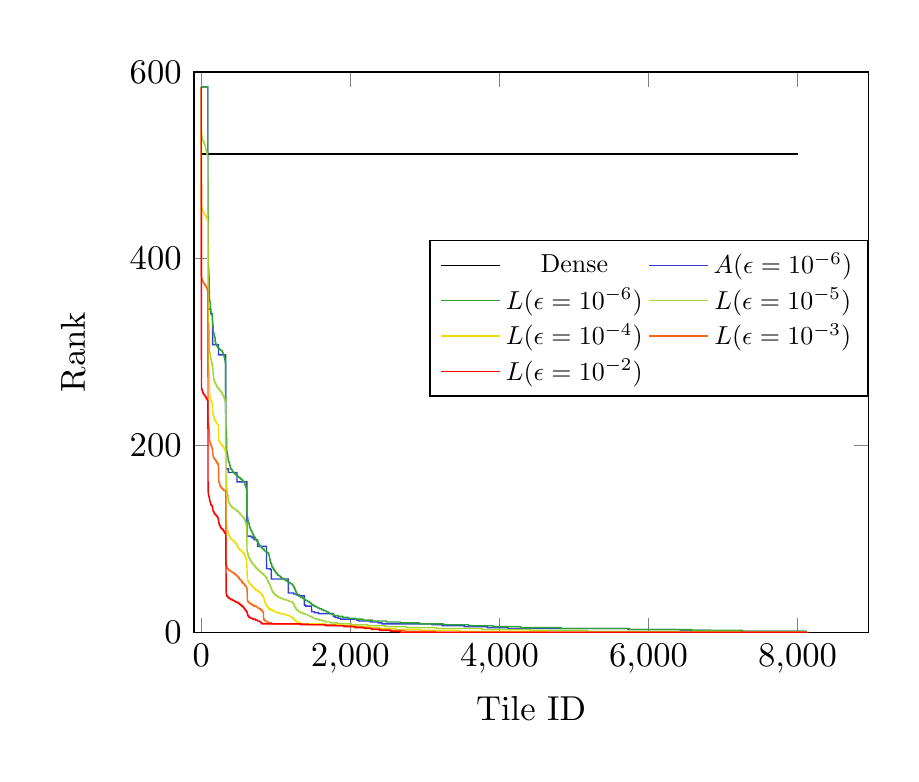}
	\caption{Rank distribution for the 3D fractional diffusion preconditioner of size $N=2^{17}$ and tile 
size of $1024$ for various compression thresholds $\epsilon$.}
	\label{fig:rank_plot_leaf_1024_3D_fd}
	\end{subfigure}
	\hfill	
	\begin{subfigure}[t]{0.49\textwidth}
	\centering
	\includegraphics[width=\textwidth]{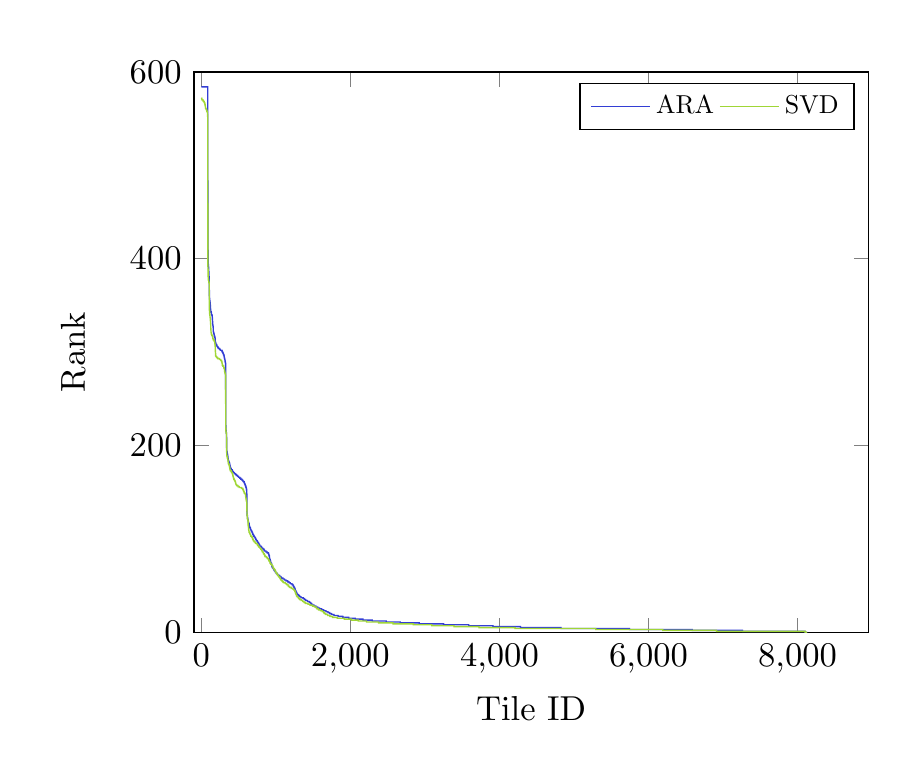}
	\caption{Difference between the ranks detected by ARA and SVD for $\epsilon=10^{-6}$.}
	\label{fig:svd_vs_ara_rank_plot_leaf_1024_3D_fd}
	\end{subfigure}
	\caption{Rank distribution for the 3D fractional diffusion Cholesky preconditioner.}
	\label{fig:rank_distribution_fd}
\end{figure}

Figure \ref{fig:rank_plot_leaf_1024_3D_fd} shows the rank distribution of each preconditioner with a 
tile size of $1024$ for various compression thresholds. As expected, the overall memory savings increase 
with looser thresholds and the memory overhead for the blocks with large ranks is negligible. 

Figure \ref{fig:svd_vs_ara_rank_plot_leaf_1024_3D_fd} shows the difference between using ARA for compression 
and the SVD to determine the lowest rank, for a threshold of $\epsilon=10^{-6}$. The ranks detected by 
the ARA are only slightly higher than those detected by the SVD with a difference of about $5\%$ on average 
for the total memory consumption. If the memory savings are desired, then for about a $20\%$ increase in solve times
it is straightforward to add a post processing step to the ARA where the SVD can be applied to each low rank tile to reduce 
the rank to the minimum.

%

\begin{figure}[t]
	\begin{subfigure}[t]{0.49\textwidth}
	\centering
	\includegraphics[width=\textwidth]{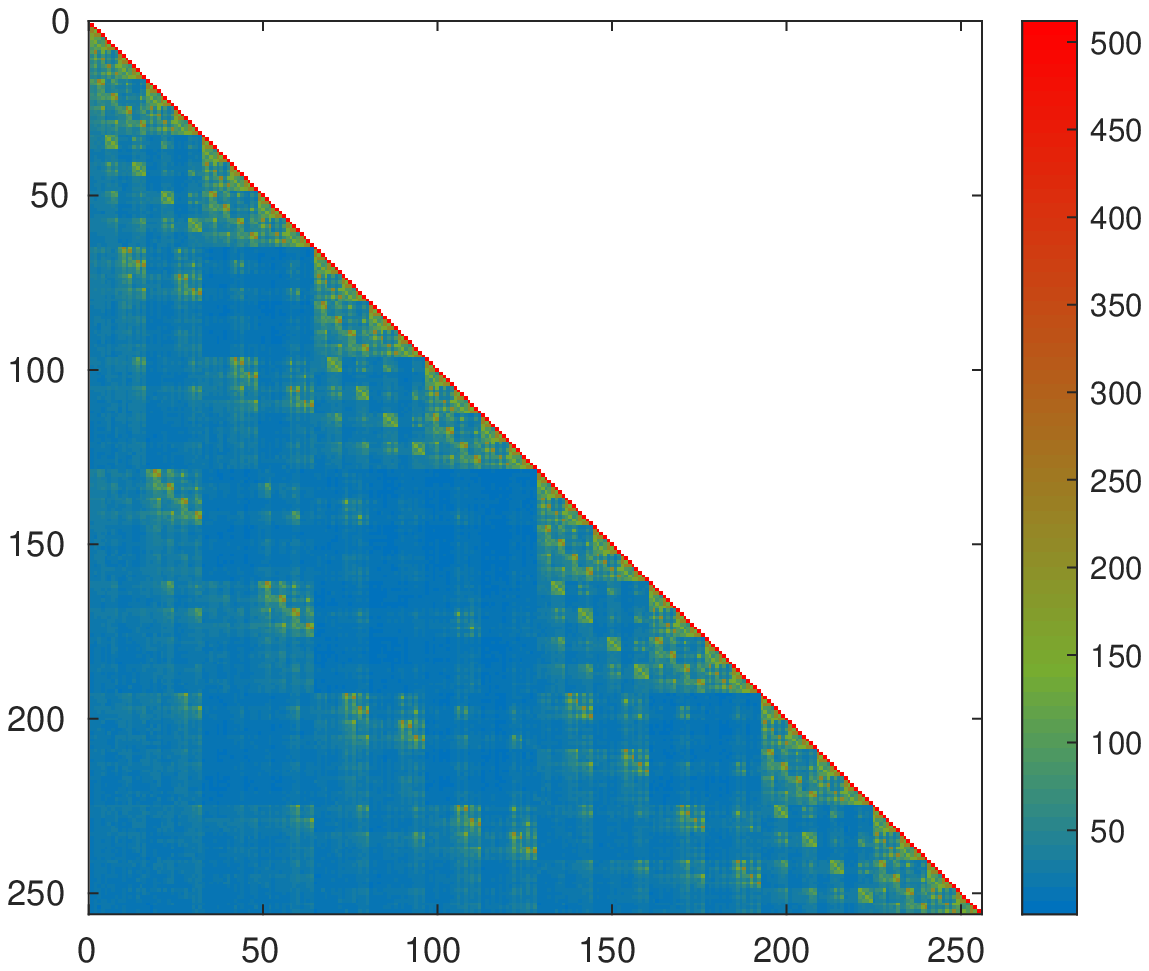}
	\caption{Heatmap of the ranks, without pivoting.}
	\label{fig:unpiv}
	\end{subfigure}
	\hfill	
	\begin{subfigure}[t]{0.49\textwidth}
	\centering
	\includegraphics[width=\textwidth]{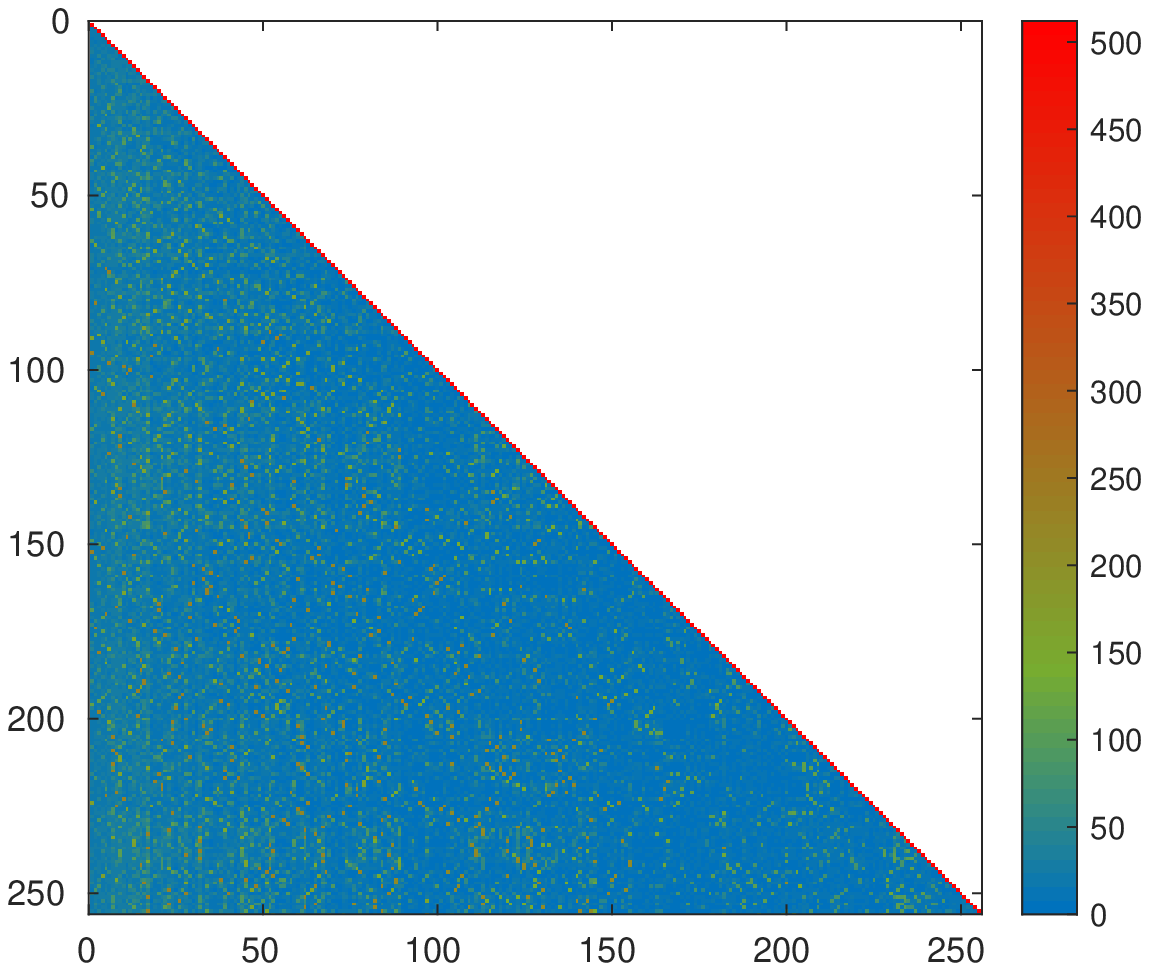}
	\caption{Heatmap of the ranks, with pivoting.}
	\label{fig:piv}
	\end{subfigure}
	\caption{Difference in tile ranks of the factorization of a $N=2^{17}$ 3D covariance matrix with a tile size of 512 and a compression threshold of $\epsilon=10^{-6}$ when pivoting is used.}
	\label{fig:pivoting_heatmap}
\end{figure}

\begin{figure}
	\begin{subfigure}[t]{0.49\textwidth}
	\centering
	\includegraphics[width=\textwidth]{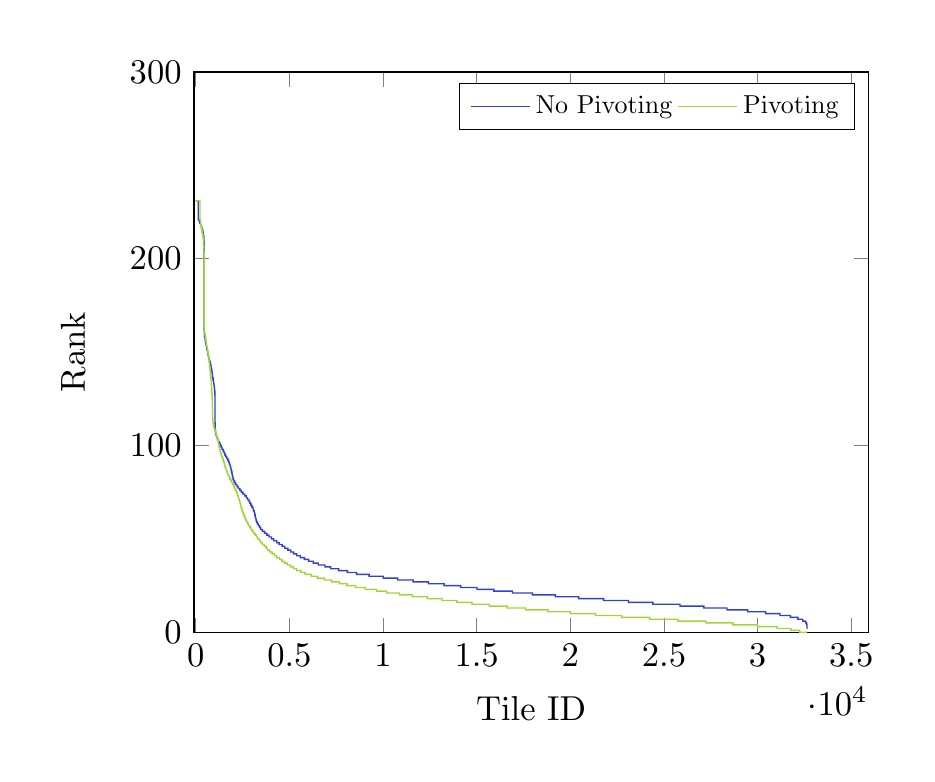}
	\caption{Rank distribution for a covariance matrix with and without pivoting showing a decrease in ranks due to pivoting.}
	\end{subfigure}
	\hfill	
	\begin{subfigure}[t]{0.49\textwidth}
	\centering
	\includegraphics[width=\textwidth]{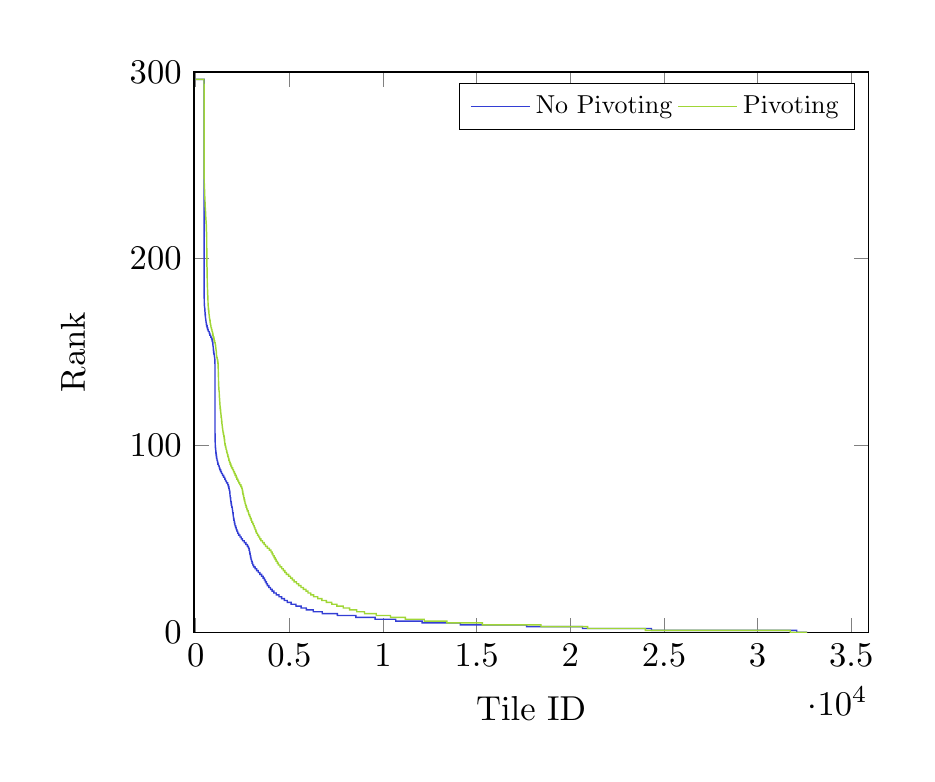}
	\caption{Rank distribution for a fractional diffusion matrix with and without pivoting showing an increase in ranks due to random choice of pivots.}
	\end{subfigure}
	\caption{Rank distribution changes due to pivoting for 3D covariance and fractional diffusion matrices of size $N=2^{17}$ and tile size of $512$.}
	\label{fig:pivoted_rank_dist}
\end{figure}

\subsection{Performance Implications of Pivoting}
\label{sec:res_extensions}
To show the effects of the extensions mentioned in Section \ref{sec:extensions} on performance, we factorize a 3D covariance TLR matrix of size $N=2^{17}$, a tile size of $512$ and a compression threshold of $\epsilon=10^{-6}$ using tile pivoted Cholesky and the $LDL^T$ factorization on the CPU. For pivoting, we test pivot selection using the approximated 2-norm as determined by the power method and the Frobenius norm. Pivot selection took about 28s using the 2-norm and 2.7s using the Frobenius norm. Interestingly, the average rank was reduced in both cases to 24 from 32 in the unpivoted case. The overall factorization time also decreases in both cases due to overall lower ranks from the 135s of the unpivoted to 128s for the 2-norm pivot selection and 103s for the Frobenius norm selection. Figure \ref{fig:pivoting_heatmap} shows the rank heatmaps for the TLR factors before and after pivoting. The ranks are significantly less clustered in the pivoted case but are overall lower than those of the unpivoted. 

To explore how ranks due to pivoting can increase and impact factorization times beyond just pivot selection, we factored a fractional diffusion matrix and chose a random pivot among those exceeding a minimum norm.  In this case factorization time increased to 159s using the 2-norm and 131s using the Froebnius norm from 101s and the average rank increased to 20 from 16. These changes in rank due to pivoting merit further investigation from a numerical analysis standpoint. Figure \ref{fig:pivoted_rank_dist} shows the effect of pivoting on the rank distribution of the triangular factors for both problems. 

Finally, the costs of the $LDL^T$ factorization are slightly lower overall than the unpivoted Cholesky with almost identical ranks at 128s. The diagonal block factorizations are almost 7x slower than the Cholesky case, but since Schur compensation is no longer necessary to maintain positive definiteness and can thus be ignored, the overall time is lower.


\section{Conclusion and Future Work}
\label{sec:conclusions}

We presented high performance TLR factorizations algorithms for symmetric matrices that are performance portable 
due to their reliance on batched matrix-matrix multiplications for the vast majority of their operations. 
The algorithm works well as a direct solver if high accuracy is requested or as a preconditioner 
when using lower accuracy approximations. 
The resulting implementation performs well 
on both CPUs and GPUs and shows the expected asymptotic growth in memory consumption and factorization 
time. 

There are a number of ways performance can be improved even further, mainly during the sampling 
process. In particular, the majority of the matrix-matrix products are with very low rank tiles that 
prevent architectures that favor high arithmetic intensity from performing as well as they 
can. One way to remedy this would be to store the low rank factors belonging to the same block row in 
a way that allows multiple products to be carried out as a single large product. Using mixed precision 
for the storage of the TLR matrix is another way of improving performance in the sampling, as offdiagonal 
tiles could be stored in a lower precision than the diagonal blocks while still sampling in the higher 
precision. This would be particularly useful on recent GPUs that are equipped with specialized tensor 
cores that can provide significant performance boosts to mixed precision computations. More advanced scheduling could also help improve occupancy by starting the sampling process for tiles that are dynamically 
determined to be ready for compression. Autotuning the tile size and parameters like the number of parallel 
buffers would also provide some desirable usability improvements.

Other TLR algorithms like the LU decomposition,
matrix-matrix products, and inversion can be implemented using the ARA framework developed in this work,
and they will be the subject of future investigations in the context of domain decomposition preconditioners \cite{zampini2016robustness,zampini2016pcbddc}
and Hessian matrices of optimization problems \cite{ambartsumyan2020hierarchical}.




\end{document}